\documentclass[
reprint,
superscriptaddress,
amsmath,amssymb,
aps,
notitlepage
]{revtex4-1}

\usepackage{color}

\usepackage{graphicx}
\usepackage{dcolumn}
\usepackage{bm}
\usepackage{physics}%
\usepackage{multirow}
\usepackage[table,xcdraw]{xcolor}
\usepackage{colortbl}

\usepackage{hyperref}

\newcommand{\Div}{$\mathcal{S}^{(n)}_{d}$}
\newcommand{\Con}{$\mathcal{S}^{(n)}_{c}$}
\newcommand{\Exc}{$\mathcal{S}^{(n)}_{e}$}
\newcommand{\Excinf}{$\mathcal{S}_{e}$}
\newcommand{\Log}{$\mathcal{S}_{l}$}
\newcommand{\Regime}{critical }

\usepackage{braket}
\usepackage{mathtools}
\usepackage{comment}

\begin{document}

\title{Fine structure of the nonlinear Drude weights in the spin-1/2 XXZ chain}

\begin{abstract}
We study nonlinear Drude weights (NLDWs) for the spin-1/2 XXZ chain in the critical regime at zero temperature. The NLDWs are generalizations of the linear Drude weight. Via the nonlinear extension of the Kohn formula, they can be read off from higher-order finite-size corrections to the ground-state energy in the presence of a $U(1)$ magnetic flux. The analysis of the ground-state energy based on the Bethe ansatz reveals that the NLDWs exhibit convergence, power-law, and logarithmic divergence, depending on the anisotropy parameter $\Delta$. We determine the convergent and power-law divergent regions, which depend on the order of the response $n$. Then, we examine the behavior of the NLDWs at the boundary between the two regions and find that they converge for $n=0, 1, 2$ $({\rm mod}~4)$, while they show logarithmic divergence for $n=3$ $({\rm mod}~4)$. Furthermore, we identify particular anisotropies $\Delta=\cos(\pi r/(r+1))$ ($r=1,2, 3,\ldots$) at which the NLDW at any order $n$ converges to a finite value.
\end{abstract}

\author{Yuhi Tanikawa}
\affiliation{%
Department of Physics, The University of Tokyo, 7-3-1 Hongo, Bunkyo-ku, Tokyo 113-0033, Japan}%

\author{Hosho Katsura}
\affiliation{%
Department of Physics, The University of Tokyo, 7-3-1 Hongo, Bunkyo-ku, Tokyo 113-0033, Japan}%
\affiliation{%
Institute for Physics of Intelligence, The University of Tokyo, 7-3-1 Hongo, Bunkyo-ku, Tokyo 113-0033, Japan}%
\affiliation{%
Trans-scale Quantum Science Institute, University of Tokyo, Bunkyo-ku, Tokyo 113-0033, Japan}%

\date{\today}

\maketitle

\section{Introduction}

Transport phenomena have been a subject of central interest in condensed matter physics.
In particular, anomalous transport properties of one-dimensional quantum many-body systems have been actively investigated since they are quite different from higher dimensional ones~\cite{zotos2005,zotos2004transport,Bertini2020,Sirker2020,hirobe2017one}.
Nevertheless, our theoretical understanding of them was rather limited to the linear response regime~\cite{Kubo} or non-interacting systems.
Thus, the theoretical study of nonlinear transport in strongly interacting systems is highly challenging.
More recently, nonlinear Drude weight (NLDW) characterizing the nonlinear static transport has been introduced~\cite{Watanabe-Oshikawa, Watanabe-Oshikawa-Liu}.
This quantity is a straightforward extension of the linear Drude weight first proposed by Kohn~\cite{Kohn} as an indicator to distinguish between a conductor and an insulator in quantum many-body systems.
Given that the linear one has played an essential role in characterizing linear transport properties~\cite{Fye1991,Stafford1993,fujimoto2003,Kirchner1999,Sirker2009,Urichuk2021}, we expect that its nonlinear counterparts will be equally or even more important in understanding transport phenomena.

There are already a number of previous studies focusing on the NLDWs~\cite{tanikawa2021exact,fava2021hydrodynamic,takasan2021adiabatic,fukusumi2021kubos}.
In Ref.~{\cite{tanikawa2021exact}}, the NLDWs in the spin-1/2 XXZ chain, which is a paradigmatic example of a quantum many-body system, was examined in detail.
It was found that they diverge in certain anisotropy parameter regimes in the thermodynamic limit.
In addition, the origin of these divergences was identified as \textit{nonanalytic} finite-size corrections to the ground-state energy~{\cite{tanikawa2021exact}}.
However, this property was discussed except when the anisotropy parameter takes special values, and thus there are still some cases that have not been thoroughly investigated.
Therefore, further research on the NLDWs in this fundamental model is needed to achieve a complete understanding.

In this paper, we investigate the fine structure of the NLDWs at zero temperature for the spin-1/2 XXZ chain in the whole \Regime regime.
The advantage of this model is its solvability by the Bethe ansatz~\cite{Takahashi, Korepin_book}.
Since the NLDWs can be read off from the higher-order finite-size corrections to the ground-state energy in the presence of a $U(1)$ flux, it is essential to analyze these corrections in detail.
This is achieved by using the Bethe ansatz, in conjunction with a mathematical method called the Wiener-Hopf method~\cite{yang1966two,Hamer, Takahashi, Sirker, Morse,tanikawa2021exact}.
Furthermore, since the Bethe ansatz enables us to treat very large
systems numerically, we can confirm the asymptotic behaviors of the NLDWs in the large system-size limit.
From the perspective of the Wiener-Hopf method, we reveal that the finite-size scaling of the ground state energy is quite distinct depending on the value of the anisotropy parameter.

The two main findings of this study are as follows.
The first one is the behaviors of the NLDWs at their boundaries between the convergent and divergent regions.
The detailed analysis suggests that the $n$th order one there converges for $n=0,1,2$ $({\rm mod}~4)$, while it shows logarithmic divergence for $n=3$ $({\rm mod}~4)$ in the large system-size limit.
By using the exact solutions, we calculate the first several orders of the NLDWs numerically and confirm their behaviors around the boundaries.
The other one is the existence of particular anisotropies where all the NLDWs converge.
Since higher order ones have  
wider divergent regions, some of the special anisotropies are surrounded by the divergent region.
We confirm this discontinuous behavior in the {\Regime} regime by calculating one of the higher order NLDWs numerically.

Our paper is organized as follows: In Sec.~\ref{introNLDW}, we review the Bethe ansatz for the XXZ chain with the $U(1)$ flux and introduce the nonlinear Kohn formula to calculate the NLDWs.
In Sec.~\ref{overview}, the main results of our study are summarized.
In Sec.~\ref{dcregions}, we review the origin of the divergences of the NLDWs and carefully determine the convergent and divergent regions.
In Sec.~\ref{boundaryp}, by considering logarithmic corrections to the ground state energy, we analytically identify the behaviors of the NLDWs at their boundaries between the convergent and divergent regions and also confirm them numerically.
In Sec.~\ref{exceptionalp}, we analytically and numerically reveal that there exist some exceptional points where all the NLDWs converge.
Finally, the discussion and conclusion of our paper are presented in Sec.~\ref{Conslusion}.
In Appendices, we provide the derivation of the finite-size scaling of the ground-state energy based on the Wiener-Hopf method.
Furthermore, numerical confirmation of the scaling for several anisotropies is also given there.

\begin{figure}[htbp]
    \includegraphics[width=0.9\hsize]{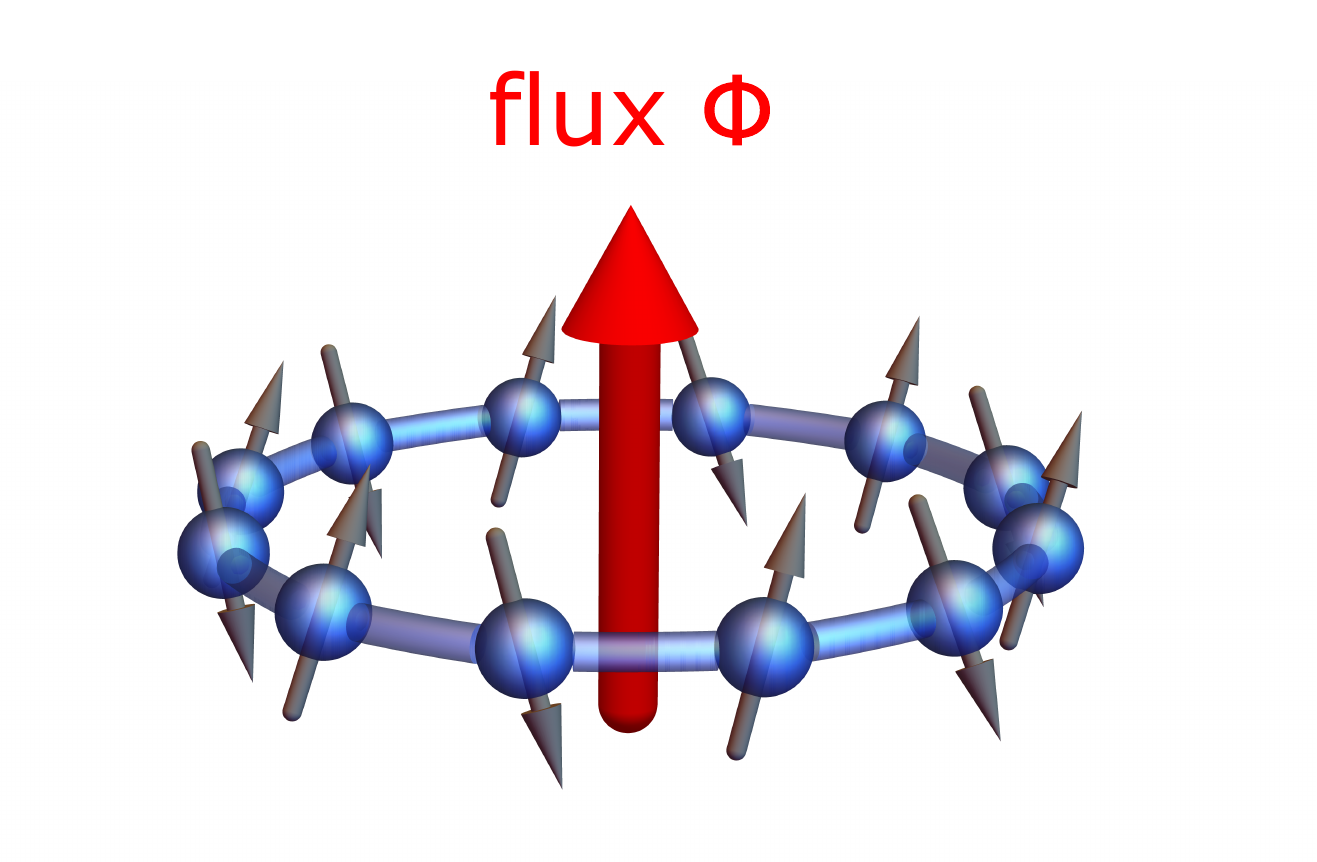}
    \caption{A one-dimensional spin chain with the $U(1)$ flux $\Phi$.}
    \label{spinchain}
\end{figure}

\begin{figure}[htbp]
    \includegraphics[width=\hsize]{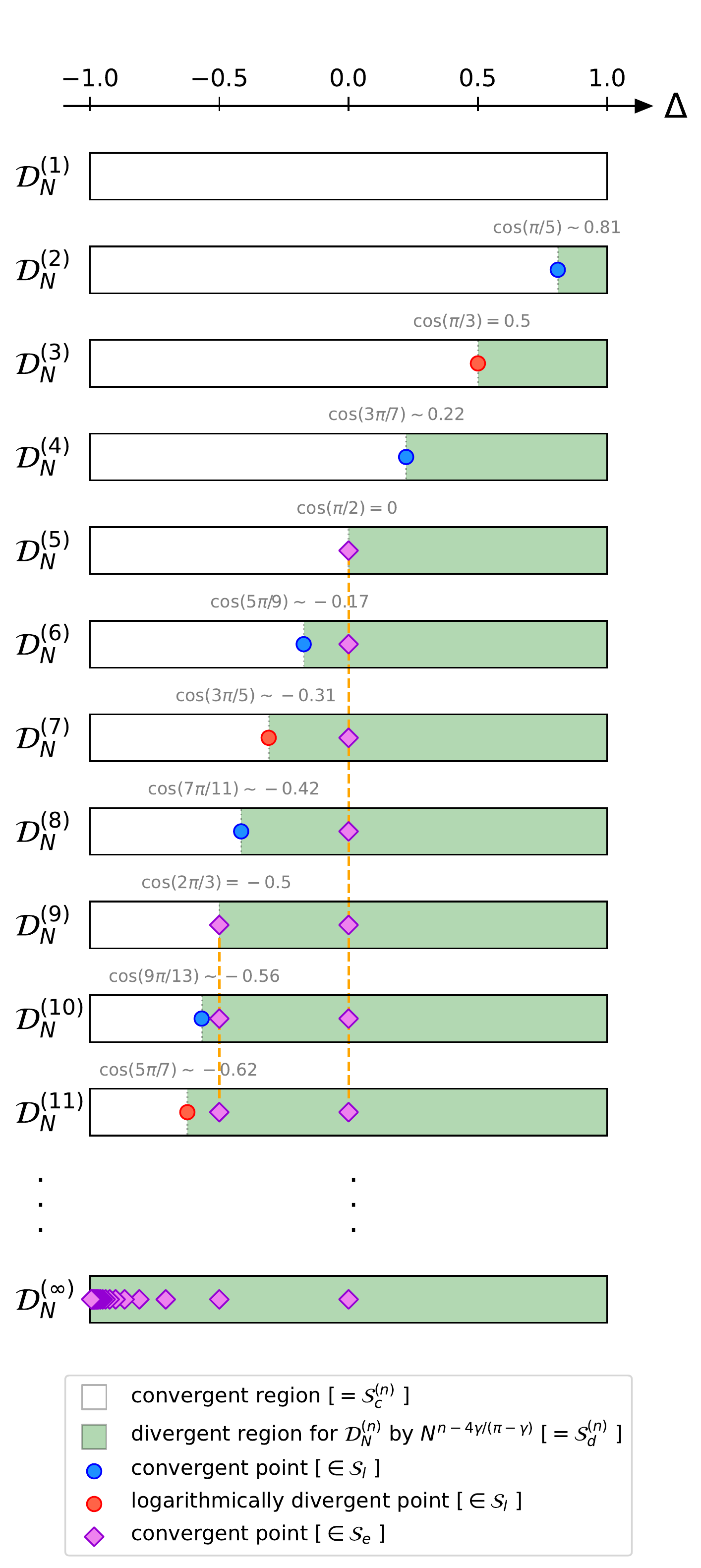}
    \caption{Fine structure of the NLDWs throughout the \Regime regime.
    Clearly, higher order NLDWs have wider divergent regions.
    Also, behaviors of the NLDWs at their boundaries between the convergent (white) and divergent (green) regions  are shown.
    The boundary between them for $\mathcal{D}^{(n)}_{N}(\Theta)$ is given by $\Delta^{(n)}_{\rm B}=\cos{\gamma^{(n)}_{\rm B}}$, where $\gamma^{(n)}_{\rm B}=\pi(n-1)/(n+3)$ $(n\in\mathbb{N})$.
    Note that we here denote a set of points where the NLDWs contain logarithmic corrections as {\Log}.
    Furthermore, we can see that there exist exceptional points {\Excinf}$\big(\!\equiv\mathcal{S}^{(\infty)}_{e}\big)$ where all the NLDWs converge.}
    \label{schematic}
\end{figure}

\section{Nonlinear Drude weights in the XXZ chain}
\label{introNLDW}
We consider the spin-1/2 XXZ chain with the $U(1)$ flux $\Phi$ defined by the Hamiltonian:
  \begin{align}
    \label{H_DM}
    \hat{\mathcal{H}}(\Phi)
    &\!=\!\sum_{l=1}^{N}2J\bigg[\frac{1}{2}e^{i\frac{\Phi}{N}}\hat{S}_{l}^{+}\hat{S}_{l+1}^{-}\!\!+{\!\rm h.c.\!}+\!\Delta \hat{S}_{l}^{z}\hat{S}_{l+1}^{z}\!\bigg],\!
  \end{align}
where $\hat{S}_{l}^{\alpha}$ $(\alpha=x,y,z)$ are spin-1/2 operators, $\hat{S}^\pm_l = \hat{S}^x_l \pm i \hat{S}^y_l$, $J>0$ is the coupling constant, $\Delta$ is the anisotropy parameter, and $N$ is the number of sites. We impose periodic boundary conditions on the chain and assume that $-1<\Delta<1$ and $N$ is even throughout this paper.
See Fig.~{\ref{spinchain}} for a schematic picture of our model.
Here it is enough to consider only $-\pi<\Phi\leq \pi$, as ${\cal H}(\Phi)$ and ${\cal H}(\Phi+2\pi)$ have the same spectrum.
We note in passing that the case with $\Phi\neq0$ corresponds to the spin-1/2 XXZ chain with the Dzyaloshinskii-Moriya (DM) interaction with a uniform DM vector along the $z$ axis~\cite{Alcaraz1990}.

{\renewcommand\arraystretch{1.7}
\setlength{\arrayrulewidth}{1.1pt}
\begin{table*}[htb]
  \caption{The leading order of $\mathcal{D}^{(n)}_N(\Theta)$ in the \Regime regime.
  The power-law divergent and logarithmic divergent behaviors in the large-$N$ limit are shown in {\bf bold} characters.
  Here we denote the boundary between the convergent and divergent regions as $\Delta_{\rm B}^{(n)}\equiv\cos{\qty(\pi(n-1)/(n+3))}$.
  Note that $\mathcal{O}(x)$ and $o(x)$ are Landau symbols indicating $\mathcal{O}(x)/x\rightarrow {\rm (const.)}$ and $o(x)/x\rightarrow 0$ $(N\rightarrow\infty)$, respectively.}
  \label{table-group-all}

{
\begin{tabular}{cccccc}
\hline\hline
\multicolumn{2}{c}{\multirow{2}{*}{~order of response $n$ (mod $4$)~}} & {$~~-1<\Delta<\Delta_{\rm B}^{(n)}~~$} & $~~\Delta=\Delta_{\rm B}^{(n)}~~$ & \multicolumn{2}{c}{{$~~\Delta_{\rm B}^{(n)}<\Delta<1~~$}} \\ \cline{5-6}
\multicolumn{2}{c}{} & convergent & { boundary } & { divergent } & { exceptional } \\ \hline
\multirow{2}{*}{~odd order~} & $n=1$ & $\mathcal{O}\qty(1)$ & { $\mathcal{O}(1)$ } & \boldmath{$~~\mathcal{O}\qty(N^{n-1-\frac{4\gamma}{\pi-\gamma}})~~$} & { $\mathcal{O}\qty(1)$ } \rule[-3mm]{0mm}{10mm}\\
 & $n=3$ & $\mathcal{O}\qty(1)$ & \boldmath{{$~~~~\mathcal{O}\big(\log{N}\big)~~~~$}} & \boldmath{$~~\mathcal{O}\qty(N^{n-1-\frac{4\gamma}{\pi-\gamma}})~~$} & { $\mathcal{O}\qty(1)$ } \rule[-4mm]{0mm}{4mm}\\
~even order~ & $n=0,2$ & { $o\qty(1)$ } & $\mathcal{O}\qty(1)$ & \boldmath{$~~\mathcal{O}\qty(N^{n-1-\frac{4\gamma}{\pi-\gamma}})~~$} & { $o\qty(1)$ } \rule[-4mm]{0mm}{10mm}\\ \hline\hline
\end{tabular}
}

\end{table*}
}

Since the total magnetization $\hat{S}^{z}_{\rm tot}=\sum_{l=1}^N\hat{S}^{z}_{l}$ is conserved in this model, we can obtain the lowest energy state in each sector individually by the Bethe ansatz~\cite{Yang-Yang}. In the sector with $M$ down spins, the Bethe roots $\{v_j(\Phi)\}$ are determined by the following Bethe equation for $j=1,2,\ldots,M$:
  \begin{align}
    &p_1\big(v_j\qty(\Phi)\big)+\frac{\Phi}{N}-\frac{1}{N}\sum_{k=1}^{M}{p_2\big(v-v_k\qty(\Phi)\big)} \nonumber \\
    &\qquad \qquad \qquad \qquad \qquad=\frac{\pi}{N}\left(-M+2j-1\right), \label{eq:Bethe_eq}
  \end{align}
where $p_n(v)\equiv 2\tan^{-1}{\left(\frac{\tanh\frac{\gamma}{2}v}{\tan\frac{n\gamma}{2}}\right)}$ and $\gamma\equiv\arccos{\Delta}$. In terms of the Bethe roots, the energy density is given by
\begin{align}
    \label{EGS}
    e(\Phi; M)
    &=\frac{1}{N}\sum_{j=1}^{M}\frac{2J\sin^2{\gamma}}{\cos{\gamma}-\cosh{(\gamma {v_{j}(\Phi))}}}+\frac{J\Delta}{2}.
\end{align}
If $\Phi=0$, it is known that the ground state lies in the sector of $M=N/2$~\cite{Affleck}. Thus, for sufficiently small $\Phi$ the ground state energy density of $\mathcal{H}(\Phi)$ is $e_\mathrm{gs}(\Phi)=e(\Phi; M=N/2)$.

Nonlinear Drude weight (NLDW) is a straightforward extension of the linear Drude weight~\cite{Kohn} and can be calculated by using the nonlinear Kohn formula~\cite{Watanabe-Oshikawa, Watanabe-Oshikawa-Liu}.
At zero temperature, the $n$th order one $\mathcal{D}^{(n)}_N(\Theta)$ can be obtained as
\begin{gather}
    \label{eq:Kohn_formula}
    \mathcal{D}^{(n)}_N(\Theta)
    =N^{n+1}\pdv[n+1]{\Phi}e_{\rm gs}(\Phi)\Bigr|_{\Phi=\Theta},
\end{gather}
where $-\pi<\Theta\leq \pi$.
Note that the finite $\Theta$ corresponds to the DM interaction as mentioned above.
In the $\Theta=0$ case, only the odd orders are nonvanishing.
This is because the ground-state energy density $e_{\rm gs}(\Phi)$ is an even function of $\Phi$, which can be seen from $\hat{U}^\dagger \hat{\mathcal {H}}(\Phi) \hat{U} =  \hat{\mathcal {H}}(-\Phi)$ with the unitary operator $\hat{U}=\prod_{l=1}^{N}2\hat{S}_{l}^{x}$.

\section{Overview of the results}
\label{overview}
Here we summarize the main results of our paper.
They are shown in Fig.~\ref{schematic} and Table~\ref{table-group-all}.
As shown in Ref.~\cite{tanikawa2021exact}, the $n$th order Drude weight $\mathcal{D}_N^{(n)}(\Theta)$ has both the convergent and divergent regions.
The boundary between them is given by
\begin{gather}
 \label{defDeltaB}
 \Delta_{\rm B}^{(n)}\equiv
 \cos{\frac{\pi(n-1)}{n+3}}.
\end{gather}
In the following, we denote the convergent region as {\Con}$\ \equiv\big(\!\!-\!\!1,\,\Delta_{\rm B}^{(n)}\big)$ and the divergent one as {\Div}$\ \equiv\big(\Delta_{\rm B}^{(n)},\,1\big)\setminus\,${\Exc}, where {\Exc}$\ \equiv\{\cos{\qty(\pi r/(r+1))}~|~r=1,2,\ldots,\lfloor{(n-2)/4} \rfloor\}$ is the set of exceptional points at which the $n$th order one converges.
Note that $\lfloor x\rfloor$ is the floor function.
The definitions of frequently used symbols are summarized in Table~\ref{set_def}.

The results for the odd order NLDWs are shown in the first and second lines of Table~\ref{table-group-all}.
In the convergent region {\Con}, i.e., $-1<\Delta<\Delta_{\rm B}^{(n)}$, they converge to finite values in the thermodynamic limit.
At the boundary point {$\Delta=\Delta_{\rm B}^{(n)}$}, they show two distinct behaviors depending on the order of the response.
When $n=1~({\rm mod}~4)$, the NLDWs at their boundaries converge to finite values in the thermodynamic limit.
On the other hand, when $n=3~({\rm mod}~4)$, the large-$N$ asymptotic behavior of the NLDWs at their boundaries is the logarithmic divergence.
In the divergent region {\Div}, the large-$N$ asymptotic behavior of the NLDWs is the power-law divergence of the form $\mathcal{D}_N^{(n)}(\Theta)\sim N^{n-1-\frac{4\gamma}{\pi-\gamma}}$.
At the points in {\Exc}, the odd order NLDWs converge to finite values unlike the divergent behaviors around the points.
This behavior is the same as that in the convergent region {\Con}.

The results for the even order NLDWs are shown in the third line of Table~\ref{table-group-all}.
In the convergent region {\Con}, i.e., $-1<\Delta<\Delta_{\rm B}^{(n)}$, they vanish in the thermodynamic limit.
At the boundary point {$\Delta=\Delta_{\rm B}^{(n)}$}, they converge to finite values in the thermodynamic limit.
These values can be calculated analytically [see Eq.~(\ref{Yk})].
In the divergent region {\Div}, the large-$N$ asymptotic behavior of the NLDWs is the power-law divergence of the form $\mathcal{D}_N^{(n)}(\Theta)\sim N^{n-1-\frac{4\gamma}{\pi-\gamma}}$, as in the results for the odd order ones.
At the points in {\Exc}, the even order NLDWs vanish  unlike the divergent behaviors around the points.
This behavior is the same as that in the convergent region {\Con}.

The above results are summarized visually in Fig.~\ref{schematic}.
It clearly shows that higher order NLDWs have the wider divergent regions {\Div}.
One might think that the infinite-order NLDW diverges everywhere in the \Regime regime.
However, this is not the case because, at the exceptional points in $\mathcal{S}^{(\infty)}_{e}=\{\cos{\qty(\pi r/(r+1))}~|~r\in\mathbb{N}\}$, all the NLDWs show the convergence as we have discussed above.
The situation is illustrated in the bottom panel of Fig.~\ref{schematic}.

In the following sections, we derive these results analytically by using the Wiener-Hopf method.
Although there are some subtle points in this approach, we confirm our results by directly solving the Bethe ansatz equations numerically.

{\renewcommand\arraystretch{1.5}
\begin{table}[htb]
  \caption{Definitions of the sets. The boundary between {\Con} and {\Div} is given by $\Delta_{\rm B}^{(n)}\equiv\cos{\qty(\pi(n-1)/(n+3))}$.}
  \label{set_def}
  \begin{tabular}{wc{2cm}wc{6cm}} \hline\hline
  {Set} & {Definition}
  \\\hline
     {\Con} & $\big(-1,\,\Delta_{\rm B}^{(n)}\big)$ \rule[-3mm]{0mm}{8mm}
     \\
     {\Exc} & $\Big\{\cos{\qty(\frac{\pi r}{r+1})}~\Bigr|~r=1,2,\ldots,\lfloor{(n-2)/4} \rfloor\Big\}$ \rule[-4mm]{0mm}{9mm}
     \\
     {\Div} & $\big(\,\Delta_{\rm B}^{(n)},\,1\,\big)\setminus\,${\Exc}\rule[-4mm]{0mm}{9mm}
     \\
     {\Excinf}$\big(\equiv\mathcal{S}^{(\infty)}_{e}\big)$ & $\Big\{\cos{\qty(\frac{\pi r}{r+1})}~\Bigr|~r\in\mathbb{N}\Big\}$ \rule[-4mm]{0mm}{9mm}
     \\
     {\Log} & $\Big\{\cos{\qty(\frac{\pi(2p-1)}{2p-1+2q})}~\Bigr|~p,q\in\mathbb{N}\Big\}$ \rule[-4mm]{0mm}{9mm}
     \\ \hline\hline
  \end{tabular}
\end{table}
}

\begin{figure*}[htbp]
    \includegraphics[width=\hsize]{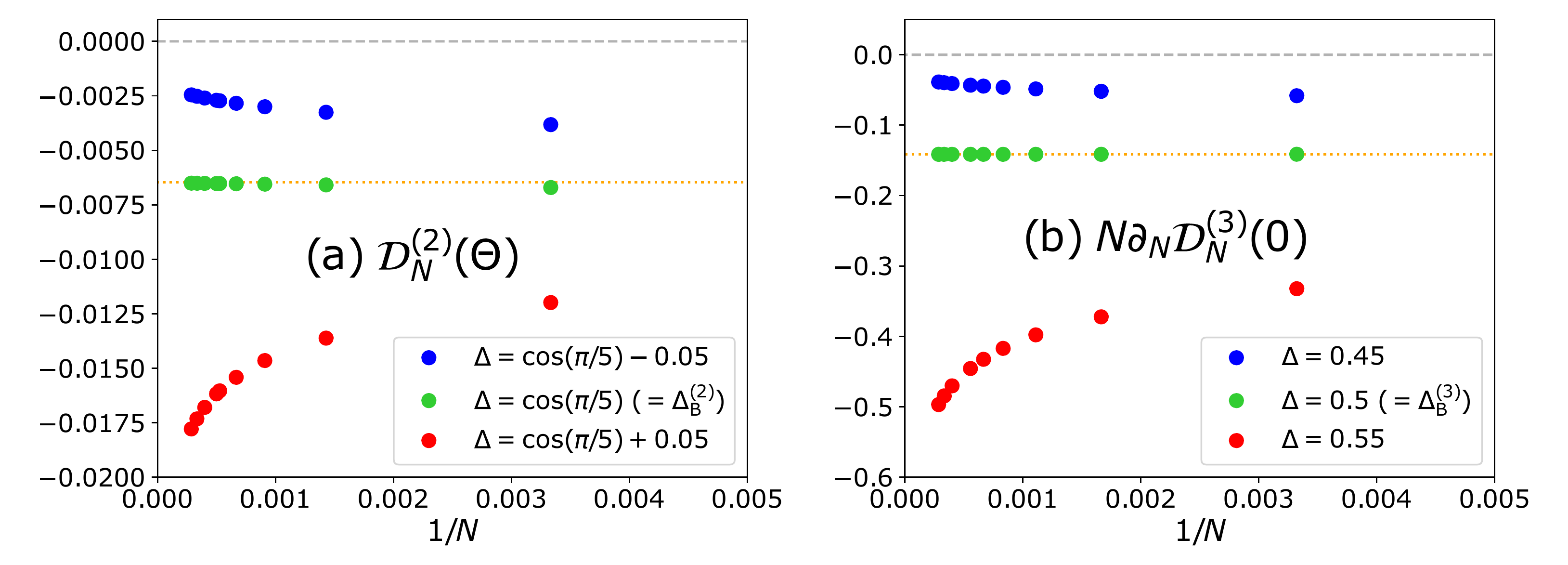}
    \caption{Numerical results for $\mathcal{D}_{N}^{(2)}(\Theta=0.1)$ around $\Delta=\Delta_{\rm B}^{(2)}=\cos{(\pi/5)}$ and $N\partial_{N}\mathcal{D}_{N}^{(3)}(0)$ around $\Delta=\Delta_{\rm B}^{(3)}=0.5$ are shown in (a) and (b), respectively.
    They have been studied for system sizes ranging from $N=300$ up to $N=3500$.
    All the vertical axes are scaled with $J$.
    Figures~\ref{D2_dD3}(a) and (b) show that both quantities vanish below the boundary (blue marker) and diverge (red marker) above the boundary.
    On the other hand, at the boundary (green marker), $\mathcal{D}_{N}^{(2)}(\Theta=0.1)$ and $N\partial_{N}\mathcal{D}_{N}^{(3)}(0)$ converge to finite values, and these values can be estimated by the linear extrapolation as  $-0.006476...$ and $-0.1414...$, respectively.
    The orange dotted lines indicate the analytical value of $\mathcal{D}_{N}^{(2)}(\Theta=0.1)$ at $\Delta=\Delta_{\rm B}^{(2)}$ and $N\partial_{N}\mathcal{D}_{N}^{(3)}(0)$ at $\Delta=\Delta_{\rm B}^{(3)}$ in the thermodynamic limit $(N\rightarrow \infty)$.}
    \label{D2_dD3}
\end{figure*}

\section{The convergent and divergent regions}
\label{dcregions}
In this section, we focus on the behaviors of the NLDWs in the convergent and divergent regions.
These regions include 
points where $e_{\rm gs}(\Phi)$ contains logarithmic corrections~[Eq.~(\ref{egs-Phi3})]. Here we 
denote the set of these points by {\Log}$\ \equiv\{\cos{\qty(\pi (2p-1)/(2p-1+2q))}~|~p,q\in\mathbb{N}\}$.
For later convenience, we also introduce the set of exceptional points {\Excinf}$\equiv\mathcal{S}^{(\infty)}_{e}\!=\{\cos{\qty(\pi r/(r+1))}~|~r\in\mathbb{N}\}$, at which all the NLDWs converge.
For $\Delta\in(-1,1)\setminus\big(${\Excinf}\,$\cup $\,{\Log}$\big)$~\footnote{Since in Ref.~\cite{tanikawa2021exact} all the exponents $4m\gamma/(\pi-\gamma)~(m\in\mathbb{N})$ are supposed to be noninteger, the points included in {\Excinf} or {\Log} are automatically excluded.}, the detailed analysis of their behaviors was given in our previous paper~\cite{tanikawa2021exact}.
From the low-energy effective field theory of the XXZ chain or the Wiener-Hopf method, the finite-size scaling of $e_{\rm gs}(\Phi)$ can be cast into the form
  \begin{align}
    \label{egs-Phi1}
    \nonumber
    e_{\rm gs}(\Phi)-e_{\rm gs}(0)=\sum_{k\geq l\geq1}A_{k,l}&\qty(\frac{1}{N})^{2k}\Phi^{2l}
    \\
    +\sum_{k,l,m\geq1}&B_{k,l,m}\qty(\frac{1}{N})^{2k+\frac{4m\gamma}{\pi-\gamma}}\Phi^{2l},
  \end{align}
where $\Delta\in(-1,1)\setminus\big(${\Excinf}\,$\cup $\,{\Log}$\big)$, and $A_{k,l}$ and $B_{k,l,m}$ are coefficients depending on $\gamma$ (see Appendix A).
Note that the smallest exponent of $1/N$ in the second sum of Eq.~(\ref{egs-Phi1}), namely $2+4\gamma/(\pi-\gamma)$, is always noninteger.
In other words, there exist nonanalytic finite-size corrections to the ground-state energy.

The straightforward differentiation of Eq.~(\ref{egs-Phi1}) with respect to $\Phi$ enables us to identify the large-$N$ asymptotic behaviors of the NLDWs.
They read
\begin{align}
    \nonumber
    &\mathcal{D}^{(2k-1)}_N(\Theta)=(2k)!\bigg[A_{k,k}+B_{1,k,1}N^{2k-2-\frac{4\gamma}{\pi-\gamma}}
    \\\label{D2n-1}
    &\qquad\qquad +\!\qty(\!A_{k+1,k}+\frac{(2k+2)!}{2(2k)!}A_{k+1,k+1}\Theta^2\!)N^{-2}\!+\cdots\bigg],
    \\\nonumber
    &\mathcal{D}^{(2k)}_N(\Theta)
    =(2k+2)!A_{k+1,k+1}\frac{\Theta}{N}
    \\\label{D2n}
    &\qquad\qquad\qquad\qquad+X_{k}(\Theta)N^{2k-1-\frac{4\gamma}{\pi-\gamma}}+\cdots,
  \end{align}
where $X_{k}(\Theta)\equiv\sum_{l> k}(2l)!/(2l-2k-1)!B_{1,l,1}\Theta^{2l-2k-1}$
~\footnote{The result for the linear Drude weight $(k=1)$ is consistent with Eq.~(20) in Ref.~\cite{laflorencie2001}, with the identification $K^*=\pi/(2 (\pi-\gamma))$}.
From these results, we can see that, in the thermodynamic limit, the odd order NLDWs converge to finite values and the even order ones vanish when $n<1+4\gamma/(\pi-\gamma)$, i.e., $-1<\Delta<\Delta_{\rm B}^{(n)}$ where $\Delta_{\rm B}^{(n)}$ is defined in Eq.~(\ref{defDeltaB}).
On the other hand, in the large-$N$ limit, $\mathcal{D}^{(n)}_N(\Theta)$ shows the power-law divergence of the form $\mathcal{D}_N^{(n)}(\Theta)\sim N^{n-1-\frac{4\gamma}{\pi-\gamma}}$ when $n>1+4\gamma/(\pi-\gamma)$, i.e., $\Delta_{\rm B}^{(n)}<\Delta<1$.

In fact, these convergent and divergent behaviors can be seen throughout the regions {\Con} and {\Div}, respectively.
To confirm this, we have to consider the remaining two cases: $\Delta\in$ {\Con}$\,\cap\,${\Excinf} and $\Delta\in\big(${\Con}$\,\cup\,${\Div}$\big)\cap$ {\Log}.
In the case $\Delta\in$ {\Con}$\,\cap\,${\Excinf}, all the coefficients $B_{k,l,m}$ in Eq.~(\ref{egs-Phi1}) vanish identically as we will see in Sec.~\ref{exceptionalp}.
This leads to the fact that the odd order NLDWs still converge to finite values and the even order ones still vanish in the thermodynamic limit.
In the other case $\Delta\in\big(${\Con}$\,\cup\,${\Div}$\big)\cap$ {\Log}, the finite-size scaling of $e_{\rm gs}(\Phi)$ contains logarithmic corrections.
However, even in these cases, when $-1<\Delta<\Delta_{\rm B}^{(n)}$, the odd order NLDWs still converge to finite values and the even order ones still vanish in the thermodynamic limit as we will see in Appendix B.
Also, when $\Delta_{\rm B}^{(n)}<\Delta<1$, the $n$th order one $\mathcal{D}^{(n)}_N(\Theta)$ still shows the power-law divergence of the form $\mathcal{D}_N^{(n)}(\Theta)\sim N^{n-1-\frac{4\gamma}{\pi-\gamma}}$ in the large-$N$ limit.

As a result, we can conclude that every NLDW $\mathcal{D}^{(n)}_N(\Theta)$ shows the convergence (or vanishing) in the convergent region {\Con} and the power-law divergence of the form $\mathcal{D}_N^{(n)}(\Theta)\sim N^{n-1-\frac{4\gamma}{\pi-\gamma}}$ in the divergent region {\Div}~(see Table~\ref{table-group-all})~\footnote{We note in passing that a divergent behavior similar to that of $\mathcal{D}^{(3)}_N(0)$ was found for the fourth derivative of the ground state energy density with respect to the magnetization~\cite{Nomura}.}.

\section{Boundary between the convergent and divergent regions}
\label{boundaryp}
As we have discussed in the previous section, the boundary point between the convergent and divergent regions of $\mathcal{D}^{(n)}_N(\Theta)$ is given by $\Delta_{\rm B}^{(n)}=\cos{\gamma^{(n)}_{{\rm B}}}$ with $\gamma^{(n)}_{{\rm B}}=\pi(n-1)/(n+3)$.
This suggests that, when $n=1~({\rm mod}~4)$, i.e., $n=4k+1~(k\in\mathbb{N})$, the boundary $\Delta^{(4k+1)}_{\rm B}=\cos{(\pi k/(k+1))}$ is included in the set of exceptional points {\Excinf}.
Since the special properties of the NLDWs at these points are discussed in Sec.~\ref{exceptionalp}, here we focus on the remaining cases: $n=0,2,3~({\rm mod}~4)$, i.e., $n=2k,4k-1~(k\in\mathbb{N})$.

The boundaries of $\mathcal{D}^{(2k)}_N(\Theta)$ and $\mathcal{D}^{(4k-1)}_N(\Theta)$ are included in {\Log} $=\{\cos{\qty(\pi (2p-1)/(2p-1+2q))}~|~p,q\in\mathbb{N}\}$ because $\Delta^{(2k)}_{{\rm B}}$ and $\Delta^{(4k-1)}_{{\rm B}}$ correspond to the cases when $p=k,\,q=2$ and $p=k,\,q=1$, respectively.
In these cases, the detailed analysis in Appendix A shows that the finite-size scaling of $e_{\rm gs}(\Phi)$ obeys
  \begin{align}
    \nonumber
    e_{\rm gs}(\Phi)-e_{\rm gs}(0)
    =\hspace{-0.7cm}\sum_{\substack{k\geq1,s\geq0\\k+s(2p-1)\geq l\geq1}}\hspace{-0.6cm}C_{k,l,s}\qty(\frac{1}{N})^{2k+2s(2p-1)}\hspace{-0.5cm}\big(\log{N}\big)^{s}\Phi^{2l}
    \\\label{egs-Phi3}
    +\hspace{-0.1cm}\sum_{\substack{k,l,m\geq1\\s\geq0}}\hspace{-0.1cm}D_{k,l,m,s}\qty(\frac{1}{N})^{2k+\frac{4m\gamma}{\pi-\gamma}+2s(2p-1)}\hspace{-0.5cm}\big(\log{N}\big)^{s}\Phi^{2l},
  \end{align}
where $\Delta\in$ {\Log}, and $C_{k,l,s}$ and $D_{k,l,m,s}$ are coefficients depending on $\gamma$ (see Appendix A).
From Eq.~(\ref{egs-Phi3}), the NLDWs at their boundaries can be calculated as
  \begin{align}
    \label{converge}
    &\mathcal{D}^{(2k)}_{N,{\rm B}}(\Theta)
    =
    \begin{cases} Y_k(\Theta)+\mathcal{O}\qty(\frac{\log{N}}{N})  & {\rm if}\ k=1,
    \\Y_k(\Theta)+\mathcal{O}\qty(\frac{1}{N}) & {\rm if}\ k\geq 2, \end{cases}
    \\\label{log}
    &\mathcal{D}^{(4k-1)}_{N,{\rm B}}(\Theta)
    =(4k)!C_{1,2k,1}\log{N}+\mathcal{O}\qty(1),
  \end{align}
where $Y_{k}(\Theta)\equiv\sum_{l> k}(2l)!/(2l-2k-1)!D_{1,l,1,0}\Theta^{2l-2k-1}$.
Note that the subscript ``B" is introduced to indicate the value at the boundary.
The above results mean that $\mathcal{D}^{(2k)}_{N,{\rm B}}(\Theta)$ converges to the finite value $Y_{k}(\Theta)$ in the thermodynamic limit.
On the other hand, $\mathcal{D}^{(4k-1)}_{N,{\rm B}}(\Theta)$ shows the logarithmic divergence in the large-$N$ limit (see Table~\ref{table-group-all}).
The analytical form of $Y_{k}(\Theta)$ can be obtained from Eq.~(4.1) in Ref.~\cite{Lukyanov} as
\begin{align}
  \nonumber
  &Y_{k}(\Theta)=
  -\frac{16\pi J\sin{\gamma}\sin{\big(\frac{2\pi\gamma}{\pi-\gamma}\big)}}{\gamma}
  \cdot
  \frac{\Gamma^2{\big(\frac{\pi}{\pi-\gamma}\big)}\Gamma^2{\big(-\frac{2\pi}{\pi-\gamma}\big)}}
  {\Gamma^2{\big(-\frac{\pi}{\pi-\gamma}\big)}}
  \\\label{Yk}
  &\quad\quad
  \Bigg[
  \frac{(\pi-\gamma)\,\Gamma{\big(\frac{\pi-\gamma}{2\gamma}\big)}}{\sqrt{\pi}\,\Gamma{\big(\frac{\pi}{2\gamma}\big)}}
  \Bigg]^{\frac{4\gamma}{\pi-\gamma}}
  \!\!
  \dv[2k+1]{\Theta}
  \frac{\Gamma{\big(\frac{\Theta+2\pi}{2(\pi-\gamma)}\big)}\Gamma{\big(\frac{-\Theta+2\pi}{2(\pi-\gamma)}\big)}}{\Gamma{\big(\frac{\Theta-2\gamma}{2(\pi-\gamma)}\big)}\Gamma{\big(\frac{-\Theta-2\gamma}{2(\pi-\gamma)}\big)}}.
\end{align}

These behaviors can be confirmed numerically.
The numerical results for $\mathcal{D}_{N}^{(2)}(\Theta=0.1)$ and $N\partial_{N}\mathcal{D}_{N}^{(3)}(0)$ around their boundaries are shown in Fig.~{\ref{D2_dD3}}(a) and (b), respectively.
Figure~{\ref{D2_dD3}}(a) shows that, in the large-$N$ region, the data points for $\mathcal{D}_{N}^{(2)}(\Theta=0.1)$ at the boundary fall almost on a straight line and approaching the analytical value indicated by the orange dotted line.
From Fig.~{\ref{D2_dD3}}(a), the value of $\mathcal{D}_{N,{\rm B}}^{(2)}(\Theta=0.1)/J$ at $1/N=0$ can be estimated as $-0.006476...$ by the linear extrapolation.
Although this value is slightly different from the analytical value $Y_{1}(\Theta=0.1)/J= -0.006470...$ at $\gamma=\gamma^{(2)}_{{\rm B}}=\pi/5$, we believe that this difference is due to numerical errors in the finite-differentiation and the extrapolation process.
We can also see that, in the large-$N$ limit, the behavior of $\mathcal{D}_{N}^{(2)}(\Theta=0.1)$ above and below the boundary becomes diverging and vanishing, respectively, which is consistent with Eq.~(\ref{D2n}).

Figure~{\ref{D2_dD3}}(b) shows that the data for $N\partial_{N}\mathcal{D}^{(3)}_{N}(0)$ at the boundary 
fall almost on a straight line in the large-$N$ region.
Note that by calculating the quantities related to the derivative of the NLDWs with respect to $N$, we can avoid observing directly the logarithmic divergence of the NLDWs themselves, which is very difficult to identify numerically.
As a result, we can confirm that $\mathcal{D}_{N}^{(3)}(0)$ shows the logarithmic divergence at the boundary as we have expected from Eq.~(\ref{log}).
This is because Eq.~(\ref{log}) yields
  \begin{gather}
      N\partial_{N}\mathcal{D}^{(4k-1)}_{N,{\rm B}}(\Theta)
    =(4k)!C_{1,2k,1}+o\qty(1),
  \end{gather}
and thus, for the case $k=1$, we have $N\partial_{N}\mathcal{D}^{(3)}_{N,{\rm B}}(0)=4!C_{1,2,1}+o\qty(1).$
Here we have approximated the derivative with respect to $N$ by finite differences.
From Fig.~{\ref{D2_dD3}}(b), the value of $N\partial_{N}\mathcal{D}^{(3)}_{N,{\rm B}}(0)/J$ at $1/N=0$ can be estimated as $-0.1414...$ by the linear extrapolation.
On the other hand, the analytical expression for $4!C_{1,2,1}$ can be obtained as $-81\sqrt{3}J/(32\pi^3)$ by setting $\gamma=\pi/3+\epsilon$ and expanding Eq.~(4.1) of Ref.~\cite{Lukyanov} in $\epsilon$ around $\epsilon=0$.
Thus we can indeed confirm that $N\partial_{N}\mathcal{D}^{(3)}_{N,{\rm B}}(0)$ in Fig.~{\ref{D2_dD3}}(b) converges to the analytical value $4!C_{1,2,1}/J=-81\sqrt{3}/(32\pi^3)=-0.1414...$ indicated by the orange dotted line.
We can also see that the behavior of $N\partial_{N}\mathcal{D}_{N}^{(3)}(0)$ above and below the boundary becomes diverging and vanishing in the large-$N$ limit, respectively.
The same holds for general $k$ and can be understood from the following relation
  \begin{align}
      \nonumber
      N\partial_{N}\mathcal{D}^{(4k-1)}_N&(0)=(4k)!\bigg[-2A_{2k+1,2k}N^{-2}
      \\\label{Ndiv}
      +&\qty(4k-2-\frac{4\gamma}{\pi-\gamma})B_{1,2k,1}N^{4k-2-\frac{4\gamma}{\pi-\gamma}}+\cdots\bigg],
  \end{align}
which follows from Eq.~(\ref{D2n-1}).
Since the power of the second term in Eq.~(\ref{Ndiv}) is the same as one appearing in Eq.~(\ref{D2n-1}) for $\mathcal{D}^{(4k-1)}_N(\Theta)$, the above quantity in the convergent region $\mathcal{S}^{(4k-1)}_{c}$ vanishes in the thermodynamic limit.
On the other hand, the large-$N$ asymptotic behavior of the above quantity is the power-law divergence of the form $N\partial_{N}\mathcal{D}^{(4k-1)}_N(0)\sim N^{4k-2-4\gamma/(\pi-\gamma)}$ in the divergent region $\mathcal{S}^{(4k-1)}_{d}$.

\begin{figure}[tbp]
    \includegraphics[width=\hsize]{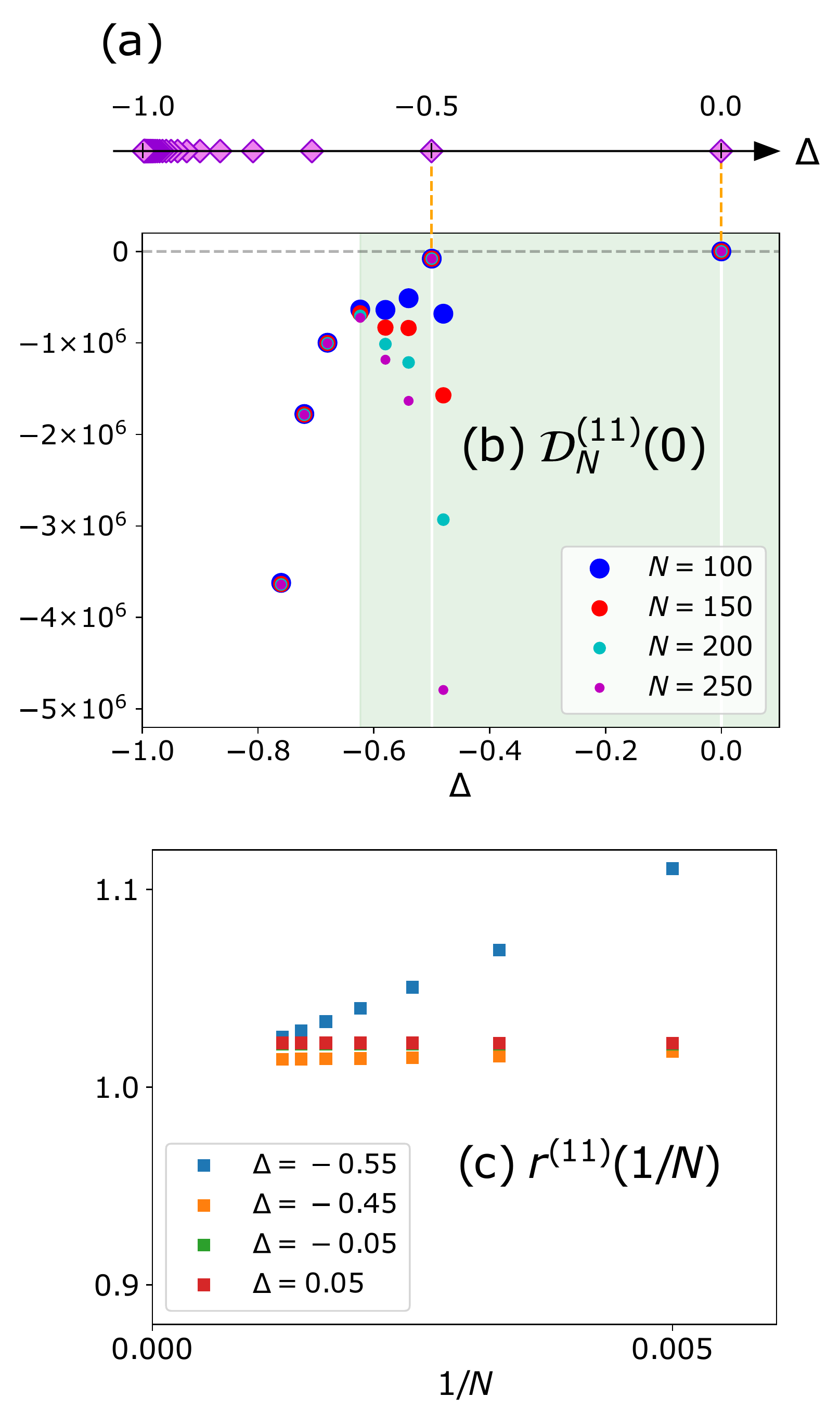}
    \caption{(a) The points in {\Excinf}$\ =\{\cos{\qty(\pi r/(r+1))}~|~r\in\mathbb{N}\}$. At these points, all the NLDWs converge in the thermodynamic limit.
    (b) The numerical result for $\mathcal{D}_{N}^{(11)}(0)$ where the vertical axis is scaled with $J$.
    The figure shows that $\mathcal{D}_{N}^{(11)}(0)$ converges to a finite value at $\Delta=0$ and $-0.5$.
    The green region represents a part of the divergent region $\mathcal{S}^{(11)}_{d}$.
    (c) The numerical results for $r^{(11)}(1/N)\equiv\mathcal{D}_{N}^{(11)}(0)/(12!\,B_{1,6,1}N^{10-4\gamma/(\pi-\gamma)})$ around $\Delta=0$ and $-0.5$.
    They indicate the presence of {\it noninteger} powers of $1/N$ in Eq.~(\ref{D2n-1}).
    }
    \label{D11}
\end{figure}

\section{Exceptional points}
\label{exceptionalp}
Now we focus on the behavior of the NLDWs at the exceptional points {\Excinf}$=\{\cos{\qty(\pi r/(r+1))}~|~r\in\mathbb{N}\}$.
These points have a special property that all the coefficients $B_{k,l,m}$ in Eq.~(\ref{egs-Phi1}) vanish identically.
This can be derived analytically by using the Wiener-Hopf method (see Appendix A).
As a result, the finite-size scaling of $e_{\rm gs}(\Phi)$ can be written as
  \begin{align}
    \label{egs-Phi2}
    e_{\rm gs}(\Phi)-e_{\rm gs}(0)=\sum_{k\geq l\geq1}A_{k,l}&\qty(\frac{1}{N})^{2k}\Phi^{2l}
  \end{align}
for $\Delta\in$ {\Excinf}.
For example, we can obtain the exact form of $e_{\rm gs}(\Phi)$ at the free-fermion point ($\Delta=0$)~\cite{Watanabe-Oshikawa,Sutherland,PhysRevB.79.245414} as
  \begin{gather}
    \label{free}
    e^{\rm free}_{\rm gs}(\Phi)-e^{\rm free}_{\rm gs}(0)=\frac{2J}{N\sin\qty(\frac{\pi}{N})}\left(1-\cos{\qty(\frac{\Phi}{N})}\right).
  \end{gather}
The large-$N$ expansion of Eq.~(\ref{free}) consists of the terms $\Phi^{\alpha}/N^{\beta}$ with $\alpha$, $\beta$ integers and $\alpha \le \beta$. Thus, nonanalytic finite-size corrections do not appear in the expansion.
This is obviously consistent with Eq.(\ref{egs-Phi2}).

Interestingly, in the thermodynamic limit, all the NLDWs converge at any $\Delta$ in {\Excinf} (see Fig.~\ref{D11}~(a)).
This can be seen by noting that
  \begin{align}
     \label{(ii)2n-1}
     &\mathcal{D}^{(2k-1)}_N(\Theta)=(2k)!A_{k,k}+\mathcal{O}\qty(\frac{1}{N^2}),
     \\\label{(ii)2n}
     &\mathcal{D}^{(2k)}_N(\Theta)=(2k+2)!A_{k+1,k+1}\frac{\Theta}{N}+\mathcal{O}\qty(\frac{1}{N^3}),
  \end{align}
which means that the odd order NLDWs remain finite, while the even order ones vanish in the thermodynamic limit (see Table~\ref{table-group-all}).
These convergent behaviors of all the NLDWs are consistent with the prediction based on numerical studies of small systems~\cite{liu2021bloch}.

We can confirm these behaviors by numerically solving the Bethe ansatz equations for large system sizes.
The exceptional points where all the NLDWs converge are shown in Fig.~\ref{D11}(a). There are infinitely many such points and they accumulate at the ferromagnetic point $\Delta=-1$.
The numerical result for $\mathcal{D}_{N}^{(11)}(0)$ is shown in Fig.~\ref{D11}(b).
As we can see in Fig.~\ref{D11}(b), the two points $\Delta=0,\,-0.5$ included in $\mathcal{S}^{(11)}_{e}$ are surrounded by the divergent region $\mathcal{S}^{(11)}_{d}$ which is colored in green.
However, $\mathcal{D}_{N}^{(11)}(0)$ at these points converge unlike the divergent behaviors around the points.
Since the divergent behaviors there should be caused by {\it noninteger} power terms of $N$ in Eq.~(\ref{D2n-1}), we calculated $r^{(11)}(1/N)\equiv\mathcal{D}_{N}^{(11)}(0)/(12!\,B_{1,6,1}N^{10-4\gamma/(\pi-\gamma)})$ numerically.
The result for this quantity is shown in Fig.~\ref{D11}(c).
This figure clearly shows that each data is on a straight line to the value near $1$ in the large $N$ region as we have expected.

Some remarks 
are in order. First, the spin-1/2 XXZ chain with periodic boundary conditions has a special symmetry related to the $\mathfrak{sl}_{2}$ loop algebra~\cite{deguchi2001sl2,miao2021q} at the exceptional points {\Excinf}~\footnote{We also note that the number of coupled nonlinear integral equations arising in the thermodynamic Bethe ansatz becomes finite at these points~\cite{Takahashi,takahashi-suzuki}.}.
We speculate that this symmetry is responsible for the convergence of all the NLDWs in the thermodynamic limit.
Second, the coefficient of the umklapp scattering term (the cosine term) in the low energy effective Hamiltonian of the XXZ chain vanishes at these points (see Eq.~$(2.23)$ in \cite{Lukyanov}).
Considering that this term leads to the nonanalytic finite-size corrections and gives the leading contribution to the power-law divergence, we can see that its vanishing is consistent with the convergence of all the NLDWs.
Finally, the ground state energy $e_{\rm gs}(\Phi)$ has the peculiar adiabatic period at these points.
For this case the adiabatic period of $e_{\rm gs}(\Phi)$ is of the order of the system size $N$, while for the other cases the period is $4\pi$~\cite{PhysRevB.46.14583}.
Based on this property, numerical calculations for small system sizes have recently revealed that the current density exhibits nontrivial oscillations, so-called Bloch oscillations, at the points in {\Excinf} even under an infinitesimal external field~\cite{liu2021bloch}.

\section{Discussion and conclusion}
\label{Conslusion}
In this paper, we examined the fine structure of the NLDWs at zero temperature for the spin-1/2 XXZ chain in the \Regime regime~(see Fig.~\ref{schematic}).
In order to calculate the NLDWs, we investigated the finite-size corrections to the ground-state energy of the chain with $U(1)$ flux and revealed that its finite-size scaling was quite distinct depending on the anisotropy parameter $\Delta$.
Based on the expansions~Eqs.~(\ref{egs-Phi1}), (\ref{egs-Phi3}) and (\ref{egs-Phi2}), we studied the large-size asymptotic behavior of the NLDWs both analytically and numerically.
The analysis determined the convergent and divergent regions of the NLDWs, the boundary of which depends on the order of the response $n$.
We studied the behaviors of the NLDWs at the boundaries in detail and found that they converge for $n=0,1,2$ (mod $4$), while they show the logarithmic divergence for $n=3$ (mod $4$) in the large system-size limit (see Table~\ref{table-group-all}).
In addition, we numerically confirmed not only the convergence but also the logarithmic divergence at the boundaries of the first several orders of the NLDWs (see Fig.~\ref{D2_dD3}).
Furthermore, we revealed that there exist special values of $\Delta$ where all the NLDWs converge in the thermodynamic limit.
Since higher order ones have wider divergent regions, some of the special $\Delta$ are surrounded by the divergent region.
We confirmed this discontinuous behavior in the {\Regime}regime by calculating one of the higher order NLDWs numerically (see Fig.~\ref{D11}).

In order to obtain the finite-size scaling of the ground-state energy, we employed the Wiener-Hopf method for the finite-size system, which is based on the Euler-Maclaurin formula~\cite{Spivey,GKP-Concrete-Math-2nd}.
Traditionally, when calculating the leading finite-size corrections to the ground state energy, higher order terms included in the expansion by this formula are often ignored~\cite{Woynarovich,Hamer}.
In general, there is no guarantee that these terms are negligible to calculate the corrections in other problems~\cite{eckle2019models,GRANET201896}.
Thus, in our study, we took all these higher order terms into account and obtained the higher order corrections to $e_{\rm gs}(\Phi)$ as well as the leading ones.
Here we should note that, although this enables us to overcome the above problem, we cannot determine the coefficients of these corrections in closed form within this approach.
Also, we assume that $e_{\rm gs}(\Phi)$ can be Taylor-expanded around $\Phi=0$ based on the symmetry of the model and comparison with the analytical results in the thermodynamic limit~\cite{tanikawa2021exact}.
Therefore, although we have confirmed our results numerically for several $\Delta$, a more rigorous derivation of the results using another method is desirable and would be an interesting future direction.

Finally, we discuss the implications of our results to the transport phenomena. One might think that the divergent behaviors of NLDWs imply the divergence of a total current density. However, this seems unlikely because contributions to the current density from different orders can cancel each other out. In fact, a similar situation is observed in a single-band tight-binding chain with a defect~{\cite{takasan2021adiabatic}}. Although the NLDWs of this system generally diverge with system size, real-time numerical simulation suggests that the adiabatic current density is suppressed compared to the defect-free case, in which the NLDW remains finite at any order.

\begin{acknowledgments}
We thank
Yoshiki Fukusumi, Yuan Miao, Kiyohide Nomura,Kazuaki Takasan, and Haruki Watanabe for valuable discussions. H. K. was supported in part by JSPS Grant-in-Aid for Scientific Research on Innovative Areas No. JP20H04630, JSPS KAKENHI Grant No. JP18K03445, and the Inamori Foundation.
\end{acknowledgments}

\appendix
\section{The finite-size corrections for the spin-1/2 XXZ chain}
\label{appendixA}

By using the Wiener-Hopf method, we calculate the finite-size corrections to the ground-state energy of the spin-1/2 XXZ chain with periodic boundary conditions:
  \begin{gather}
    \hat{\mathcal{H}}(0)=\sum_{l=1}^{N}2J\bigg[\hat{S}_{l}^{x}\hat{S}_{l+1}^{x}+\hat{S}_{l}^{y}\hat{S}_{l+1}^{y}+\Delta \hat{S}_{l}^{z}\hat{S}_{l+1}^{z}\bigg].
  \end{gather}
Although these finite-size corrections based on the same method had been partly discussed in Ref.~\cite{Woynarovich,Hamer}, here we expose the mathematical details and 
illustrate the derivation process for readers' convenience.
This detailed analysis also enables us to reveal that there are some cases with logarithmic finite-size corrections.
As a result, we derive the general expression of  the finite-size corrections including logarithmic ones.
Furthermore, after calculating these finite-size corrections, we introduce the $U(1)$ flux into them and obtain the finite-size scaling of $e_{\rm gs}(\Phi)$.

\subsection*{\ref{appendixA}-1.~Setup}
First, we review the Bethe ansatz and derive some important relations.
It is known that the ground-state energy of the above model can be obtained by this ansatz.
The Bethe roots $\{v_j\}$ are determined by the following Bethe equations for $j=1,2,\ldots,N/2$:
  \begin{gather}
    \label{BetheZ}
    \mathcal{Z}_N(v_j)=\frac{2\pi I_j}{N}=\frac{\pi}{N}\left(-\frac{N}{2}+2j-1\right),
  \end{gather}
where
  \begin{gather}
    \label{Z_N}
    \mathcal{Z}_N(v)\equiv p_1(v)-\frac{1}{N}\sum_{k=1}^{N/2}{p_2(v-v_k)}
  \end{gather}
with
  \begin{gather}
    p_n(v)\equiv 2\tan^{-1}{\left(\frac{\tanh\frac{\gamma}{2}v}{\tan\frac{n\gamma}{2}}\right)}.
  \end{gather}
Note that there exists a unique set of real solutions $\{v_j\}$ satisfying $-\infty\leq v_1<v_2<\ldots<v_{N/2}\leq\infty$ and $v_{j}=-v_{N/2-j+1}$.
Differentiating Eq.~(\ref{Z_N}) with respect to $v$, we get
  \begin{align}
    \label{rho_N}
    \rho_N(v)&\equiv\frac{1}{2\pi}\dv{\mathcal{Z}_N(v)}{v}=a_1(v)-\frac{1}{N}\sum_{k=1}^{N/2}{a_2(v-v_k)}
    \\
    &=\rho_N(-v),
  \end{align}
where
  \begin{align}
    a_n(v)&\equiv\frac{1}{2\pi}\dv{v}p_n(v)=\frac{\gamma}{2\pi}\frac{\sin n\gamma}{\cosh\gamma v-\cos n\gamma}
    \\
    &=a_n(-v).
  \end{align}
Then $\{v_j\}$ gives the ground-state energy density as
  \begin{gather}
    \label{e_N}
    e_{{\rm gs},N}=-\frac{2\pi A}{N}\sum_{j=1}^{N/2}a_1(v_j)+\frac{\Delta}{2}, 
  \end{gather}
where $A=2J \sin \gamma/\gamma$.
Now we introduce a new useful function $S_{N}(v)$ as
  \begin{gather}
    S_{N}(v)\equiv\frac{1}{N}\sum_{j=1}^{N/2}\delta(v-v_{j})-\rho_{N}(v).
  \end{gather}
This transforms Eq.~(\ref{rho_N}) into the following form:
  \begin{align}
    \nonumber
    \rho_N(v)&=a_1(v)-\int^{\infty}_{-\infty}\left(\frac{1}{N}\sum_{k=1}^{N/2}\delta(x-v_{j})\right){a_2(v-x)}dx
    \\\nonumber
    &=a_1(v)-\int^{\infty}_{-\infty}\rho_{N}(x){a_2(v-x)}dx
    \\\label{rhoN_int}
    &\qquad\quad\qquad\quad -\int^{\infty}_{-\infty}S_{N}(x){a_2(v-x)}dx.
  \end{align}
Here we define a Fourier transformation of a function $f(x)$ as
  \begin{gather}
    \tilde{f}(\omega)=\int^{\infty}_{-\infty}f(x)e^{i\omega x}dx.
  \end{gather}
By using Fourier transformation on both sides of Eq.~(\ref{rhoN_int}), we get
  \begin{align}
    \label{rho_tilde}
    \tilde{\rho}_{N}(\omega)=\frac{\tilde{a}_{1}(\omega)}{1+\tilde{a}_{2}(\omega)}-\tilde{S}_{N}(\omega)\frac{\tilde{a}_{2}(\omega)}{1+\tilde{a}_{2}(\omega)},
  \end{align}
where the Fourier transform of $a_n(v)$ is
  \begin{align}
    \tilde{a}_n\qty(\omega)&=\int^{\infty}_{-\infty}a_n(x)e^{i\omega x}dx=\frac{\sinh\qty(\frac{\pi}{\gamma}-n)\omega}{\sinh\frac{\pi}{\gamma}\omega}
    \\
    &=\tilde{a}_n\qty(-\omega).
  \end{align}
Then by using Fourier transformation on both sides of Eq.~(\ref{rho_tilde}), we obtain
  \begin{align}
    \label{rho_N-inf}
    \rho_N(v)=\rho_{\infty}(v)-\int^{\infty}_{-\infty}S_{N}(x)R(v-x)dx,
  \end{align}
 where $\rho_{\infty}(v)$ and $R(v)$ are defined as follows:
  \begin{align}
    \nonumber
    \rho_{\infty}(v)&\equiv\frac{1}{2\pi}\int^{\infty}_{-\infty}e^{-i\omega v}\frac{\tilde{a}_{1}(\omega)}{1+\tilde{a}_{2}(\omega)}d\omega
    \\\label{rho_inf}
    &=\frac{1}{2\pi}\int^{\infty}_{-\infty}\frac{e^{-i\omega v}}{2\cosh\omega}d\omega
    =\frac{1}{4\cosh\frac{\pi}{2}v}
    \\
    &=\rho_{\infty}(-v),
  \end{align}
  \begin{align}
    \nonumber
    R(v)&\equiv\frac{1}{2\pi}\int^{\infty}_{-\infty}e^{-i\omega v}\frac{\tilde{a}_2\qty(\omega)}{1+\tilde{a}_2\qty(\omega)}d\omega
    \\\label{R}
    &=\frac{1}{2\pi}\int^{\infty}_{-\infty}e^{-i\omega v}\frac{\sinh\qty(\frac{\pi}{\gamma}-2)\omega}{2\cosh\omega\sinh\qty(\frac{\pi}{\gamma}-1)\omega}d\omega
    \\
    &=R(-v).
  \end{align}
Note that $\rho_{\infty}(v)$ is the exact representation of $\rho_{N}(v)$ in the thermodynamic limit.
Similarly, Eq.~(\ref{e_N}) leads to
  \begin{align}
    \nonumber
    e_{{\rm gs},N}&=-2\pi A\int^{\infty}_{-\infty}\left(\frac{1}{N}\sum_{k=1}^{N/2}\delta(v-v_{j})\right){a_1(v)}dv+\frac{\Delta}{2}
    \\
    \nonumber
    &=-2\pi A\int^{\infty}_{-\infty}\Big(S_{N}(v)+\rho_{N}(v)\Big){a_1(v)}dv+\frac{\Delta}{2}
    \\\nonumber
    &=-2\pi A\int^{\infty}_{-\infty}\rho_{\infty}(v){a_1(v)}dv+\frac{\Delta}{2}
    \\\nonumber
    &\quad-2\pi A\int^{\infty}_{-\infty}S_{N}(v)a_{1}(v)dv
    \\\nonumber
    &\quad+2\pi A\int^{\infty}_{-\infty}\qty(\int^{\infty}_{-\infty}S_{N}(x)R(v-x)dx)a_{1}(v)dv
    \\\label{egs_N-inf}
    &=e_{{\rm gs},\infty}-2\pi A\int^{\infty}_{-\infty}S_{N}(v)\rho_{\infty}(v)dv,
  \end{align}
where we introduced
  \begin{gather}
    e_{{\rm gs},\infty}\equiv-2\pi A\int^{\infty}_{-\infty}\rho_{\infty}(v){a_1(v)}dv+\frac{\Delta}{2}.
  \end{gather}
Note that the third line follows from Eq.~(\ref{rho_N-inf}) and the last from the following relation:
  \begin{align}
    \nonumber
    &\int^{\infty}_{-\infty}\qty(S_{N}(v)-\int^{\infty}_{-\infty}S_{N}(x)R(v-x)dx)a_{1}(v)dv
    \\\nonumber
    &=\int^{\infty}_{-\infty}\qty(\frac{1}{2\pi}\int^{\infty}_{-\infty}e^{-i\omega v}\frac{\tilde{S}_{N}(\omega)}{1+\tilde{a}_{2}(\omega)}d\omega)a_{1}(v)dv
    \\\nonumber
    &=\frac{1}{2\pi}\int^{\infty}_{-\infty}\tilde{S}_{N}(\omega)\frac{\tilde{a}_{1}(\omega)}{1+\tilde{a}_{2}(\omega)}d\omega
    \\\nonumber
    &=\frac{1}{2\pi}\int^{\infty}_{-\infty}\tilde{S}_{N}(\omega)\tilde{\rho}_{\infty}(\omega)d\omega
    \\
    &
    =\int^{\infty}_{-\infty}S_{N}(v)\rho_{\infty}(v)dv.
  \end{align}
Since we can see that only the second term in Eq.~(\ref{egs_N-inf}) is responsible for the finite-size corrections to the ground-state energy, we only have to evaluate the effect of $S_{N}(v)$ to achieve the goal.

Next, we introduce a useful formula to treat $S_{N}(v)$ included in the integral.
The derivation of the formula is based on the Euler-Maclaurin formula~\cite{Spivey,GKP-Concrete-Math-2nd}:
  \begin{align}
    \nonumber
    \sum_{j=m}^{n}f(j)=\int^{n}_{m}f(x)dx&+\frac{f(m)+f(n)}{2}
    \\
    +&\int^{n}_{m}f^{\prime}(x)B_{1}(x-\lfloor x\rfloor)dx,
  \end{align}
where $f(x)$ is a continuous function, $\lfloor x\rfloor$ is the floor function, and $B_{k}(x)$ is the {\it k}th Bernoulli polynomial satisfying
  \begin{gather}
    B_{0}(x)=1
    \\\label{recurrence}
    B^{\prime}_{k}(x)=kB_{k-1}(x)\ (k\geq1)
    \\
    \int^{1}_{0}B_{k}(x)dx=0\ (k\geq1).
  \end{gather}
By using the recurrence relation (\ref{recurrence}) and integral by parts, we naively obtain
  \begin{align}
    \nonumber
    \sum_{j=m}^{n}f(j)=\int^{n}_{m}&f(x)dx+\frac{f(m)+f(n)}{2}
    \\
    +&\sum^{\infty}_{k=1}\frac{B_{k+1}}{(k+1)!}\qty(f^{(k)}(n)-f^{(k)}(m)),
  \end{align}
where $B_{k}=B_{k}(0)$ is the $k$th Bernoulli number.
Although we have $B_{2l+1}=0~(l\in\mathbb{N})$, we keep these terms explicit in the following discussion.
The above relation and the fact that $I_{j+1}-I_{j}=1$ give us the following relation:
  \begin{align}
    &\sum_{j=1}^{N/2}f(v_{j})=\sum_{I=I_{1}}^{I_{N/2}}f\left(Z_{N}^{-1}\qty(\frac{2\pi I}{N})\right)
    \\\nonumber
    &\ \ =\int^{I_{N/2}}_{I_1}f\left(Z_{N}^{-1}\qty(\frac{2\pi x}{N})\right)dx
    +\frac{f(v_{1})+f(v_{N/2})}{2}
    \\
    &\quad\quad+\sum_{k=1}^{\infty}\frac{B_{k+1}}{(k+1)!}\dv[k]{x}f\left(Z_{N}^{-1}\qty(\frac{2\pi x}{N})\right)\Biggr|^{x=I_{N/2}}_{x=I_1}
    \\\nonumber
    &\ \ =N\int^{v_{N/2}}_{v_1}f(v)\rho_{N}(v)dv+\frac{f(v_{1})+f(v_{N/2})}{2}
    \\
    &\ \ \ \quad+\sum_{k=1}^{\infty}\frac{B_{k+1}}{N^{k}(k+1)!}\qty(\frac{1}{\rho_{N}(v)}\frac{\rm d}{{\rm d}v})^{k}f(v)\Biggr|^{v=v_{N/2}}_{v=v_1},
  \end{align}
where we defined $v(x)\equiv {Z}_{N}^{-1}(2\pi x/N)$ satisfying $v(I_{j})={Z}_{N}^{-1}(2\pi I_{j}/N)=v_{j}$ and used
  \begin{align}
    Z_{N}\qty(v(x))=\frac{2\pi x}{N}\Rightarrow\dv{Z_{N}}{v}=2\pi\rho_{N}(v)=\frac{2\pi}{N}\dv{x}{v}.
  \end{align}
Therefore, we obtain
  \begin{align}
    \nonumber
    &\int^{\infty}_{-\infty}S_{N}(v)f(v)dv
    =\frac{1}{N}\sum_{j=1}^{N/2}f(v_j)-\int^{\infty}_{-\infty}f(v)\rho_{N}(v)dv
    \\\nonumber
    &\ =-\qty(\int^{\infty}_{v_{N/2}}+\int^{v_1}_{-\infty})f(v)\rho_{N}(v)dv
    +\frac{f(v_{1})+f(v_{N/2})}{2N}
    \\\label{SN_int}
    &\quad\quad +\frac{1}{N}\sum_{k=1}^{\infty}\frac{B_{k+1}}{N^{k}(k+1)!}\qty(\frac{1}{\rho_{N}(v)}\frac{\rm d}{{\rm d}v})^{k}f(v)\Biggr|^{v=v_{N/2}}_{v=v_1}.
  \end{align}
It is obvious that the above relation enables us to evaluate the finite-size corrections in Eqs.~(\ref{rho_N-inf})~and~(\ref{egs_N-inf}).

Finally, we introduce important relations employed in the Wiener-Hopf method briefly.
In the following discussion, we denote $v_{N/2}(=-v_{1})$ as $\Lambda$.
By using Eq.~(\ref{SN_int}), we get
  \begin{align}
    \nonumber
    &\rho_{N}(v)-\rho_{\infty}(v)
    \\\nonumber
    &\ =\qty(\int^{\infty}_{\Lambda}+\int^{-\Lambda}_{-\infty})R(v-u)\rho_{N}(u)du
    \\\nonumber
    &\quad -\frac{R(v+\Lambda)+R(v-\Lambda)}{2N}
    \\\label{rho_finite}
    &\quad -\frac{1}{N}\sum_{k=1}^{\infty}\frac{B_{k+1}}{N^{k}(k+1)!}\qty(\frac{1}{\rho_{N}(u)}\frac{\rm d}{{\rm d}u})^{k}R\qty(v-u)\Biggr|^{u=\Lambda}_{u=-\Lambda}
    \\
    \nonumber
    &e_{{\rm gs},N}-e_{{\rm gs},\infty}
    \\\nonumber
    &\ =2\pi A~\Biggl\{
    \qty(\int^{\infty}_{\Lambda}+\int^{-\Lambda}_{-\infty})\rho_{\infty}(v)\rho_{N}(v)dv
    \\\nonumber
    &\quad-\frac{\rho_{\infty}(\Lambda)+\rho_{\infty}(-\Lambda)}{2N}
    \\\label{egs_finite}
    &\quad-\frac{1}{N}\sum_{k=1}^{\infty}\frac{B_{k+1}}{N^{k}(k+1)!}\qty(\frac{1}{\rho_{N}(v)}\frac{\rm d}{{\rm d}v})^{k}\rho_{\infty}\qty(v)\Biggr|^{v=\Lambda}_{v=-\Lambda}\Biggr\}.
  \end{align}
These are the complete representations of the finite-size corrections using $N$ and $\Lambda$.
Thus, in order to obtain the corrections using only $N$, we have to derive the relation between $N$ and $\Lambda$.
(Actually, we can roughly identify $e^{-\qty(\pi/2)\Lambda}$ with $1/N$ as we will see in the following discussion.)
Now we introduce new functions
  \begin{align}
    g(v)\equiv\rho_{N}(v+\Lambda)=g_{+}(v)+g_{-}(v),
  \end{align}
  \begin{gather}
    \label{g}
    g_{\pm}(v)\equiv\Theta\qty(\pm v)g(v),
  \end{gather}
where $\Theta\qty(v)$ is a Heaviside step function.
Then by substituting $v+\Lambda$ to the argument of Eq.~(\ref{rho_finite}), we have
  \begin{align}
    \nonumber
    &g(v)-\rho_{\infty}(v+\Lambda)
    \\\nonumber
    &\ =\int^{\infty}_{-\infty}\Big\{R(v-u)+R(v+u+2\Lambda)\Big\}g_{+}(u)du
    \\\nonumber
    &\qquad\ \ -\frac{R(v+2\Lambda)+R(v)}{2N}
    \\\nonumber
    &\qquad\ \ -\frac{1}{N}\sum_{k=1}^{\infty}\mathcal{P}_{k}\qty(\frac{1}{N},\Big\{\rho^{(n)}_{N}(\Lambda)\Big\})R^{(k)}(v)
    \\\label{g-relation}
    &\qquad\ \ -\frac{1}{N}\sum_{k=1}^{\infty}\mathcal{Q}_{k}\qty(\frac{1}{N},\Big\{\rho^{(n)}_{N}(\Lambda)\Big\})R^{(k)}(v+2\Lambda)
  \end{align}
where we introduced coefficients $\mathcal{P}_{k}$ and $\mathcal{Q}_{k}$ depending on $1/N$ and $\rho^{(n)}_{N}(\Lambda)\qty(=g^{(n)}(0))$ for $n\geq0$, and superscripts denote numbers of derivatives.
Note that Eq.~(\ref{rho_finite}) suggests that all the terms included in $\mathcal{P}_{k}$ or $\mathcal{Q}_{k}$ can be expressed as follows:
  \begin{align}
    \label{Pk}
    {\rm (const.)}\times\frac{1}{N^l}\frac{\prod_{n=1}^{\infty}{\qty(\rho^{(n)}_{N}(\Lambda))^{N_{n}}}}{\big(\rho_{N}(\Lambda)\big)^m}
  \end{align}
where $l,m,N_{n}\in\mathbb{Z}_{\geq0}$, and each power satisfies $\sum_{n}N_n=m-l$ and $\sum_{n}{n N_n}=l-k$.
Since we have $\rho^{(n)}_{N}(\Lambda)\sim\mathcal{O}(e^{-(\pi/2)\Lambda})\sim\mathcal{O}(1/N)$, which can be seen in the following discussion, Eq.~(\ref{Pk}) implies $\mathcal{P}_{k},\mathcal{Q}_{k}\sim\mathcal{O}(1)$.

Here we investigate behaviors of $\rho_{\infty}(v+\Lambda)$ and $R(v+2\Lambda)$ for $v>0$, which appear in Eq.(\ref{g-relation}).
Since Eqs.~(\ref{rho_inf})~and~(\ref{R}) give
  \begin{align}
    \nonumber
    \rho_{\infty}(v+\Lambda)&=\frac{1}{2\pi}\int^{\infty}_{-\infty}\frac{e^{-i\omega (v+\Lambda)}}{2\cosh\omega}d\omega
    \\
    &=\frac{1}{2\pi}\int^{\infty}_{-\infty}\tilde{\rho}_{\infty}(\omega)e^{-i\omega (v+\Lambda)}d\omega,
    \\\nonumber
    R(v+2\Lambda)&=\frac{1}{2\pi}\int^{\infty}_{-\infty}\!\!e^{-i\omega (v+2\Lambda)}\frac{\sinh\qty(\frac{\pi}{\gamma}-2)\omega}{2\cosh\omega\sinh\qty(\frac{\pi}{\gamma}-1)\omega}d\omega
    \\
    &=\frac{1}{2\pi}\int^{\infty}_{-\infty}\tilde{R}(\omega)e^{-i\omega (v+2\Lambda)}d\omega,
  \end{align}
we can see that poles of $\tilde{\rho}_{\infty}(\omega)$ or $\tilde{R}(\omega)$ in the lower-half plane contribute to $\rho_{\infty}(v+\Lambda)$ and $R(v+2\Lambda)$, respectively.
The position of the poles can be read off from the explicit expressions for $\tilde{\rho}_{\infty}(\omega)$ and $\tilde{R}(\omega)$ as follows:
  \begin{gather}
    \label{poles1}
    \tilde{\rho}_{\infty}(\omega)
    \rightarrow
    {\rm poles}:\omega=-i\pi\qty(p-\frac{1}{2}),
    \\\label{poles2}
    \tilde{R}(\omega)
    \rightarrow
    {\rm poles}:\omega=-i\pi\qty(p-\frac{1}{2}), -i\frac{q\pi\gamma}{\pi-\gamma},
  \end{gather}
where $p, q\in\mathbb{N}$.
Since $\tilde{R}(\omega)$ have poles dependent on the parameter $\gamma$, in order to obtain the finite-size corrections, we must consider whether all the poles of $\tilde{R}(\omega)$ are distinct or not.
Thus, we perform the following classification shown in Table~\ref{table}.
Actually, this classification is essential to ensure convergence of coefficients $A_{k,l}, B_{k,l,m}, C_{k,l,s}$ and $D_{k,l,m,s}$ appearing in Eqs.~{(\ref{egs-finite-corr}),(\ref{exp-special}) and (\ref{finite-corr-log})}.

{\renewcommand\arraystretch{1.5}
\begin{table}[htb]
  \caption{Classification of the finite-size corrections.}
  \label{table}
  \begin{tabular}{wc{1cm}wc{6cm}} \hline\hline
    {Case} & Value of parameter $\Delta$ \\ \hline
     (i) & $(-1,1)\setminus\big(${\Excinf}\,$\cup $\,{\Log}$\big)$ \rule[-3mm]{0mm}{8mm}\\
     (ii) & $\Big\{\cos{\qty(\frac{\pi r}{r+1})}~\Bigr|~r\in\mathbb{N}\Big\}~\big(\equiv$ {\Excinf}\,$\big)$\rule[-4mm]{0mm}{9mm} \\
     (iii) & $\Big\{\cos{\qty(\frac{\pi(2p-1)}{2p-1+2q})}~\Bigr|~p,q\in\mathbb{N}\Big\}~\big(\equiv$ {\Log}\,$\big)$ \rule[-4mm]{0mm}{9mm}\\ \hline\hline
  \end{tabular}
\end{table}
}

\subsection*{\ref{appendixA}-2.~Case (i): $\Delta\in$$(-1,1)\setminus\big(${\Excinf}\,$\cup $\,{\Log}$\big)$}
Here we consider the case of $\gamma\ne\pi(2p-1)/(2p-1+2q)\ {\rm or}\ \pi r/(r+1)\ (p,q,r\in\mathbb{N})$.
In this case, all the poles of $\tilde{\rho}_{\infty}(\omega)$ $\big(\tilde{R}(\omega)\big)$ are distinct simple poles.
Thus, we have
  \begin{align}
    \nonumber
    &\rho_{\infty}(v+\Lambda)
    \\
    &=\sum_{p^{\prime}\geq1}{\rm Res}\left(\tilde{\rho}_{\infty},-i\pi\qty(p^{\prime}-\frac{1}{2})\right)\cdot\frac{e^{-\pi\qty(p^{\prime}-\frac{1}{2})(v+\Lambda)}}{i}
    \\\nonumber
    &={\rm Res}\qty(\tilde{\rho}_{\infty},-i\frac{\pi}{2})\cdot\frac{e^{-\frac{\pi}{2}(v+\Lambda)}}{i}
    \\&
    \qquad\quad+{\rm Res}\qty(\tilde{\rho}_{\infty},-i\frac{3\pi}{2})\cdot\frac{e^{-\frac{3\pi}{2}(v+\Lambda)}}{i}+\cdots,
  \end{align}
  \begin{align}
    \nonumber
    &R(v+2\Lambda)
    \\\nonumber
    &=\sum_{p^{\prime}\geq1}{\rm Res}\left(\tilde{R},-i\pi\qty(p^{\prime}-\frac{1}{2})\right)\cdot\frac{e^{-\pi\qty(p^{\prime}-\frac{1}{2})(v+2\Lambda)}}{i}
    \\\label{R(v+2lambda)}
    &\qquad\ +\sum_{q^{\prime}\geq1}{\rm Res}\left(\tilde{R},-i\frac{q^{\prime}\pi\gamma}{\pi-\gamma}\right)\cdot\frac{e^{-\frac{q^{\prime}\pi\gamma}{\pi-\gamma}(v+2\Lambda)}}{i}
    \\\nonumber
    &={\rm Res}\qty(\tilde{R},-i\frac{\pi}{2})\cdot\frac{e^{-\frac{\pi}{2}(v+2\Lambda)}}{i}
    \\&
    \quad+{\rm Res}\qty(\tilde{R},-i\frac{\pi\gamma}{\pi-\gamma})\cdot\frac{e^{-\frac{\pi\gamma}{\pi-\gamma}(v+2\Lambda)}}{i}+\cdots,
  \end{align}
for $v>0$.
Here we denoted a residue of a function $f(x)$ at $x=x_{0}$ as ${\rm Res}\qty(f,x_{0})$.
It is obvious that poles closer to the real axis contribute to the smaller power of $e^{-\qty(\pi/2)\Lambda}$.
Therefore Eq.~(\ref{g-relation}) implies that $g(v)$ can also be expanded as
  \begin{gather}
    \label{exp_g}
    g(v)=g^{[1]}(v)+g^{[2]}(v)+\cdots,
  \end{gather}
where superscripts denote increasing powers of $e^{-\qty(\pi/2)\Lambda}$ or $1/N$.
Then by substituting Eq.~(\ref{exp_g}) into Eq.~(\ref{g-relation}) and extracting the same order terms, we obtain, for examples,
  \begin{align}
    \nonumber
    g^{[1]}&(v)-\Big[\rho_{\infty}\qty(v+\Lambda)\Big]^{[1]}
    \\\nonumber
    &=\int^{\infty}_{-\infty}R\qty(v-u)g^{[1]}_{+}(u)du-\frac{R(v)}{2N}
    \\
    &\quad-\Bigg[\frac{1}{N}\sum_{k=1}^{\infty}\mathcal{P}_{k}\qty(\frac{1}{N},\Big\{\rho^{(n)}_{N}(\Lambda)\Big\})R^{(k)}(v)\Bigg]^{[1]},
  \end{align}
  \begin{align}
    \nonumber
    &g^{[2]}(v)-\Big[\rho_{\infty}\qty(v+\Lambda)\Big]^{[2]}
    \\\nonumber
    &\ \ =\int^{\infty}_{-\infty}R\qty(v-u)g^{[2]}_{+}(u)du
    \\\nonumber
    &\ \quad+\Bigg[\int^{\infty}_{-\infty}R\qty(v+u+2\Lambda)g^{[1]}_{+}(u)du\Bigg]^{[2]}
    -\Bigg[\frac{R(v+2\Lambda)}{2N}\Bigg]^{[2]}
    \\\nonumber
    &\ \quad-\Bigg[\frac{1}{N}\sum_{k=1}^{\infty}\mathcal{P}_{k}\qty(\frac{1}{N},\Big\{\rho^{(n)}_{N}(\Lambda)\Big\})R^{(k)}(v)\Bigg]^{[2]}
    \\
    &\ \quad-\Bigg[\frac{1}{N}\sum_{k=1}^{\infty}\mathcal{Q}_{k}\qty(\frac{1}{N},\Big\{\rho^{(n)}_{N}(\Lambda)\Big\})R^{(k)}(v+2\Lambda)\Bigg]^{[2]}
  \end{align}
where superscripts $[\cdots]^{[n]}$ again denote increasing powers of $e^{-\qty(\pi/2)\Lambda}$ or $1/N$.
By using Fourier transformation and integral by parts, we get
 \begin{align}
   \nonumber
   \tilde{g}^{[1]}_{+}&(\omega)+\tilde{g}^{[1]}_{-}(\omega)-\Big[\tilde{\rho}_{\infty}(\omega)e^{-i\omega \Lambda}\Big]^{[1]}
   \\\nonumber
   &=\tilde{R}(\omega)\tilde{g}^{[1]}_{+}(\omega)-\frac{\tilde{R}(\omega)}{2N}
   \\\label{g1}
   &\qquad-\Bigg[\frac{\tilde{R}(\omega)}{N}\sum_{k=1}^{\infty}\mathcal{P}_{k}\qty(\frac{1}{N},\Big\{\rho^{(n)}_{N}(\Lambda)\Big\})\qty(-i\omega)^{k}\Bigg]^{[1]}
 \end{align}
 \begin{align}
   \nonumber
   &\tilde{g}^{[2]}_{+}(\omega)+\tilde{g}^{[2]}_{-}(\omega)-\Big[\tilde{\rho}_{\infty}(\omega)e^{-i\omega \Lambda}\Big]^{[2]}
   \\\nonumber
   &=\tilde{R}(\omega)\tilde{g}^{[2]}_{+}(\omega)+\Big[\tilde{R}(\omega)\tilde{g}^{[1]}_{+}(-\omega)e^{-i2\omega \Lambda}\Big]^{[2]}
   \!-\!\Bigg[\frac{\tilde{R}(\omega)e^{-i2\omega\Lambda}}{2N}\Bigg]^{[2]}
   \\\nonumber
   &\qquad-\Bigg[\frac{\tilde{R}(\omega)}{N}\sum_{k=1}^{\infty}\mathcal{P}_{k}\qty(\frac{1}{N},\Big\{\rho^{(n)}_{N}(\Lambda)\Big\})\qty(-i\omega)^{k}\Bigg]^{[2]}
   \\
   &\qquad-\Bigg[\frac{\tilde{R}(\omega)e^{-i2\omega\Lambda}}{N}\sum_{k=1}^{\infty}\mathcal{Q}_{k}\qty(\frac{1}{N},\Big\{\rho^{(n)}_{N}(\Lambda)\Big\})\qty(-i\omega)^{k}\Bigg]^{[2]}.
 \end{align}
From the above relations, we can obtain the orders of $\tilde{g}^{[1]}_{+}(\omega)$ and $\tilde{g}^{[2]}_{+}(\omega)$ by splitting the whole into two parts: the part analytic in the upper half-plane and the other part analytic in the lower half-plane.
Now we recall that a Fourier transform $\tilde{f}(\omega)$ can be split as follows:
  \begin{gather}
    \tilde{f}(\omega)=\tilde{f}_{+}(\omega)+\tilde{f}_{-}(\omega),
  \end{gather}
where $\tilde{f}_{+}(\omega)$ and $\tilde{f}_{-}(\omega)$ are defined as
  \begin{align}
    \label{def_pm}
    \tilde{f}_{\pm}(\omega)&\equiv\pm \frac{i}{2\pi}\int_{-\infty}^{\infty}\frac{\tilde{f}({\omega}^{\prime})}{\omega-{\omega}^{\prime}\pm i0}d{\omega}^{\prime}
    \\
    &=\int^{\infty}_{-\infty}\Theta(\pm x)f(x)e^{i\omega x}dx
  \end{align}
and are analytic in the upper and lower half-plane, respectively (actually, $\tilde{g}^{[n]}_{\pm}(\omega)$ are examples).
We also introduce the following convenient factorization~\cite{yang1966two,Hamer, Takahashi, Sirker,tanikawa2021exact}:
 \begin{gather}
   \label{1-R=GG}
   1-\tilde{R}(\omega)=\frac{1}{G_{+}(\omega)G_{-}(\omega)},
 \end{gather}
where $G_{+}(\omega)$ and $G_{-}(\omega)$ are written as
 \begin{align}
   G_{+}(\omega)&=\frac{\sqrt{2\qty(\pi-\gamma)}\Gamma\qty(1-i\frac{\omega}{\gamma})}{\Gamma\qty(\frac{1}{2}-i\frac{\omega}{\pi})\Gamma\qty(1-i\omega\frac{\pi-\gamma}{\pi\gamma})}\qty(\frac{\qty(\frac{\pi}{\gamma}-1)^{\frac{\pi}{\gamma}-1}}{\qty(\frac{\pi}{\gamma})^{\frac{\pi}{\gamma}}})^{-i\frac{\omega}{\pi}}
   \\&
   =G_{-}(-\omega)
 \end{align}
and are analytic  and non-zero in the upper and lower half-plane, respectively.
Here we calculate $\tilde{g}^{[1]}_{+}(\omega)$ as an example.
By using the above methods for splitting, we can transform Eq.~(\ref{g1}) as
  \begin{align}
    \nonumber
    &\frac{\tilde{g}^{[1]}_{+}(\omega)}{G_{+}(\omega)}-\Big[G_{-}(\omega)\tilde{\rho}_{\infty}(\omega)e^{-i\omega \Lambda}\Big]^{[1]}_{+}+\left[G_{-}(\omega)\frac{\tilde{R}(\omega)}{2N}\right]^{[1]}_{+}
    \\\nonumber
    &\quad+\Bigg[G_{-}(\omega)\frac{\tilde{R}(\omega)}{N}\sum_{k=1}^{\infty}\mathcal{P}_{k}\qty(\frac{1}{N},\Big\{\rho^{(n)}_{N}(\Lambda)\Big\})\qty(-i\omega)^{k}\Bigg]^{[1]}_{+}
    \\\nonumber
    &=-G_{-}(\omega)\tilde{g}^{[1]}_{-}(\omega)+\Big[G_{-}(\omega)\tilde{\rho}_{\infty}(\omega)e^{-i\omega \Lambda}\Big]^{[1]}_{-}
    \\\nonumber
    &\quad-\left[G_{-}(\omega)\frac{\tilde{R}(\omega)}{2N}\right]^{[1]}_{-}
    \\\label{g3}
    &\quad-\Bigg[G_{-}(\omega)\frac{\tilde{R}(\omega)}{N}\sum_{k=1}^{\infty}\mathcal{P}_{k}\qty(\frac{1}{N},\Big\{\rho^{(n)}_{N}(\Lambda)\Big\})\qty(-i\omega)^{k}\Bigg]^{[1]}_{-}
    \\
    &\equiv P^{[1]}(\omega).
  \end{align}
We see that the left- and right-hand side of Eq.~(\ref{g3}) are analytic in the upper and lower half-plane, respectively.
Since both of them are analytic on the real axis, the right-hand side of Eq.~(\ref{g3}) is the analytic continuation of the left-hand side, and thus there should be the entirely analytic form $P^{[1]}(\omega)$ \cite{Hamer,Morse}.
Although the form of $P^{[1]}(\omega)$ is determined so that $\tilde{g}^{[1]}_{+}(\omega)\rightarrow0~(|\omega|\rightarrow\infty)$, we do not need the explicit form for our purposes.
As a result, we obtain
  \begin{align}
    \nonumber
    &\tilde{g}^{[1]}_{+}(\omega)
    \\\nonumber
    &\ =G_{+}(\omega)\Bigg\{\Big[G_{-}(\omega)\tilde{\rho}_{\infty}(\omega)e^{-i\omega \Lambda}\Big]^{[1]}_{+}-\left[G_{-}(\omega)\frac{\tilde{R}(\omega)}{2N}\right]^{[1]}_{+}
    \\\nonumber
    &\qquad\ -\bigg[G_{-}(\omega)\frac{\tilde{R}(\omega)}{N}\sum_{k=1}^{\infty}\mathcal{P}_{k}\qty(\frac{1}{N},\Big\{\rho^{(n)}_{N}(\Lambda)\Big\})\qty(-i\omega)^{k}\bigg]^{[1]}_{+}
    \\
    &\qquad\  +P^{[1]}(\omega)\Bigg\}
    \\\nonumber
    &\ =G_{+}(\omega)\Bigg\{\frac{\frac{i}{2}G_{+}\qty(i\frac{\pi}{2})}{\omega+i\frac{\pi}{2}}e^{-\frac{\pi}{2}\Lambda}+\frac{\tilde{R}(\omega)}{2NG_{+}(\omega)}
    \\\nonumber
    &\qquad\ -\bigg[G_{-}(\omega)\frac{\tilde{R}(\omega)}{N}\sum_{k=1}^{\infty}\mathcal{P}_{k}\qty(\frac{1}{N},\Big\{\rho^{(n)}_{N}(\Lambda)\Big\})\qty(-i\omega)^{k}\bigg]^{[1]}_{+}
    \\
    &\qquad\  +P^{[1]}(\omega)\Bigg\}
    \\
    &\ =c_1(\omega)e^{-\frac{\pi}{2}\Lambda}+c_2(\omega)\frac{1}{N}+\Bigg[c_3\qty(\omega,\frac{1}{N},\Big\{\rho^{(n)}_{N}(\Lambda)\Big\})\frac{1}{N}\Bigg]^{[1]}
  \end{align}
where $c_1$, $c_2$ and $c_3\qty(\sim\mathcal{O}(1))$ are certain coefficients.
The term with $e^{-\frac{\pi}{2}\Lambda}$ originally derives from the term
  \begin{align}
    G_{+}(\omega)\Big[G_{-}(\omega)\tilde{\rho}_{\infty}(\omega)e^{-i\omega \Lambda}\Big]^{[1]}_{+}
    =G_{+}(\omega)\frac{\frac{i}{2}G_{+}\qty(i\frac{\pi}{2})}{\omega+i\frac{\pi}{2}}e^{-\frac{\pi}{2}\Lambda},
  \end{align}
which is contributed by the simple pole of $\tilde{\rho}_{\infty}(\omega)$ closest to the real axis: $\omega=-i\pi/2$.
The higher order ones can be calculated in the same way.
However, since in that case the poles of $\tilde{R}(\omega)$ contribute to $\tilde{g}^{[n]}_{+}(\omega)$ via
  \begin{align}
    G_{+}(\omega)\Big[G_{-}(\omega)\tilde{R}(\omega)\tilde{g}_{+}(-\omega)e^{-i2\omega \Lambda}\Big]^{[n]}_{+}
  \end{align}
for example, $\tilde{g}^{[n]}_{+}(\omega)$ can contain special power terms like $(e^{-\frac{\pi}{2}\Lambda})^{4\gamma/(\pi-\gamma)}$.
Therefore we have
  \begin{align}
    \nonumber
    \tilde{g}_{+}(\omega)&=\tilde{g}^{[1]}_{+}(\omega)+\tilde{g}^{[2]}_{+}(\omega)+\tilde{g}^{[3]}_{+}(\omega)+\cdots
    \\\nonumber
    &=\sum_{\substack{k,l,m\geq0\\k+l={\rm odd}}}\mathcal{A}_{k,l,m}(\omega)\qty(\frac{1}{N})^{k}\Big(e^{-\frac{\pi}{2}\Lambda}\Big)^{l+\frac{4m\gamma}{\pi-\gamma}}
    \\\label{g+exp}
    &\qquad\qquad+\mathcal{B}\qty(\omega,\frac{1}{N},\Big\{\rho^{(n)}_{N}(\Lambda)\Big\})\frac{1}{N}
  \end{align}
where $\mathcal{A}_{k,l,m}$ and $\mathcal{B}(\sim\mathcal{O}(1))$ are certain coefficients.
Now we can obtain the relation between $\Lambda$ and $N$.
Recalling Eqs.~(\ref{BetheZ}) and (\ref{Z_N})
  \begin{align}
    Z_{N}(\infty)&-Z_{N}(v_{N/2})=\pi-\gamma-\frac{\pi-2\gamma}{2}-\frac{2\pi I_{N/2}}{N}
    =\frac{\pi}{N}
    \\
    &\qquad=2\pi\int^{\infty}_{v_{N/2}}\rho_{N}(v)dv=2\pi\tilde{g}_{+}(0),
  \end{align}
we get
  \begin{align}
    \nonumber
    \tilde{g}_{+}(0)&=\sum_{\substack{k,l,m\geq0\\k+l={\rm odd}}}\mathcal{A}_{k,l,m}(0)\qty(\frac{1}{N})^{k}\Big(e^{-\frac{\pi}{2}\Lambda}\Big)^{l+\frac{4m\gamma}{\pi-\gamma}}
    \\
    &\qquad\qquad+\mathcal{B}\qty(0,\frac{1}{N},\Big\{\rho^{(n)}_{N}(\Lambda)\Big\})\frac{1}{N}
    \\\label{g+(0)-Ninv}
    &=\frac{1}{2N}
    \\\nonumber
    \Rightarrow\frac{1}{N}&=\sum_{\substack{k,l,m\geq0\\k+l={\rm odd}}}2\mathcal{A}_{k,l,m}(0)\qty(\frac{1}{N})^{k}\Big(e^{-\frac{\pi}{2}\Lambda}\Big)^{l+\frac{4m\gamma}{\pi-\gamma}}
    \\\label{Ninv}
    &\qquad\qquad+2\mathcal{B}\qty(0,\frac{1}{N},\Big\{\rho^{(n)}_{N}(\Lambda)\Big\})\frac{1}{N}.
  \end{align}
Then by substituting Eq.~(\ref{Ninv}) into its right-hand side successively, we obtain
  \begin{align}
    \label{Ninv2}
    \frac{1}{N}=\!\!\sum_{k,l\geq0}\!\Bigg[\mathcal{C}_{k,l}+\mathcal{D}_{k,l}\bigg(e^{-\frac{\pi}{2}\Lambda},\Big\{\rho^{(n)}_{N}(\Lambda)\Big\}\bigg)\Bigg]\!\Big(e^{-\frac{\pi}{2}\Lambda}\Big)^{2k+1+\frac{4l\gamma}{\pi-\gamma}}
  \end{align}
where $\mathcal{C}_{k,l}$ and $\mathcal{D}_{k,l}(\sim\mathcal{O}(1))$ are certain coefficients.
Note that all the terms included in $\mathcal{D}_{k,l}$ are expressed as follows:
  \begin{align}
    {\rm (const.)}\times\Big(e^{-\frac{\pi}{2}\Lambda}\Big)^{s}~\frac{\prod_{n=1}^{\infty}{\qty(\rho^{(n)}_{N}(\Lambda))^{N_{n}}}}{\big(\rho_{N}(\Lambda)\big)^m}
  \end{align}
where $s,m,N_{n}\in\mathbb{Z}_{\geq0}$, and each power satisfies $\sum_{n}N_n=m-s$.
Here since we have
  \begin{align}
    \rho_{N}(\Lambda)&=g(0)=2g_{+}(0)=\frac{1}{\pi}\int^{\infty}_{-\infty}\tilde{g}_{+}(\omega)d\omega
    \\\nonumber
    &=\sum_{\substack{k,l,m\geq0\\k+l={\rm odd}}}\mathcal{E}_{k,l,m}\Bigg[\qty(\frac{1}{N})^{k}\Big(e^{-\frac{\pi}{2}\Lambda}\Big)^{l+\frac{4m\gamma}{\pi-\gamma}}\Bigg]
    \\&\qquad\qquad\quad+\mathcal{F}\qty(\frac{1}{N},\Big\{\rho^{(n^{\prime})}_{N}(\Lambda)\Big\})\frac{1}{N},
  \end{align}
$\rho^{(n)}_{N}(\Lambda)$ can be expressed as
  \begin{align}
    \rho^{(n)}_{N}(\Lambda)&=\dv[n]{\Lambda}\rho_{N}(\Lambda)
    =\frac{1}{\pi}\dv[n]{\Lambda}\int^{\infty}_{-\infty}\tilde{g}_{+}(\omega)d\omega
    \\\nonumber
    &=\sum_{\substack{k,l,m\geq0\\k+l={\rm odd}}}\mathcal{E}_{k,l,m}\dv[n]{\Lambda}\Bigg[\qty(\frac{1}{N})^{k}\Big(e^{-\frac{\pi}{2}\Lambda}\Big)^{l+\frac{4m\gamma}{\pi-\gamma}}\Bigg]
    \\\label{g(n)(0)}
    &\qquad\quad+\dv[n]{\Lambda}\Bigg[\mathcal{F}\qty(\frac{1}{N},\Big\{\rho^{(n^{\prime})}_{N}(\Lambda)\Big\})\frac{1}{N}\Bigg]
  \end{align}
where we introduced coefficients $\mathcal{E}_{k,l,m}$ and $\mathcal{F}$ as the integral values of $\mathcal{A}_{k,l,m}$ and $\mathcal{B}$, respectively.
Thus by substituting Eqs.~(\ref{Ninv2}) and (\ref{g(n)(0)}) into the right hand side of Eq.~(\ref{g(n)(0)}) successively and using the Maclaurin expansion with respect to $e^{-\frac{\pi}{2}\Lambda}~(\ll 1)$, we obtain
  \begin{align}
    \label{g(n)(0)2}
    \rho^{(n)}_{N}(\Lambda)=\sum_{k,l\geq 0}\mathcal{H}_{n,k,l}\Big(e^{-\frac{\pi}{2}\Lambda}\Big)^{2k+1+\frac{4l\gamma}{\pi-\gamma}},
  \end{align}
where $\mathcal{H}_{n,k,l}$ is a certain coefficient.
Since Eqs.~(\ref{g+exp}), (\ref{Ninv2}) and (\ref{g(n)(0)2}) give us
  \begin{align}
    \tilde{g}_{+}(\omega)=\sum_{k,l\geq 0}\mathcal{I}_{k,l}(\omega)\Big(e^{-\frac{\pi}{2}\Lambda}\Big)^{2k+1+\frac{4l\gamma}{\pi-\gamma}},
  \end{align}
we finally get the following relation between $N$ and $\Lambda$ from Eq.~(\ref{g+(0)-Ninv}):
  \begin{align}
    \label{N-lambda}
    \frac{1}{2N}&=\sum_{k,l\geq 0}\mathcal{I}_{k,l}(0)\Big(e^{-\frac{\pi}{2}\Lambda}\Big)^{2k+1+\frac{4l\gamma}{\pi-\gamma}}
    \\\label{lambda-N-sub}
    \Leftrightarrow
    e^{-\frac{\pi}{2}\Lambda}=\frac{1}{\mathcal{I}_{0,0}(0)}&\Bigg[\frac{1}{2N}-\sum_{k+l\geq 1}\mathcal{I}_{k,l}(0)\Big(e^{-\frac{\pi}{2}\Lambda}\Big)^{2k+1+\frac{4l\gamma}{\pi-\gamma}}\Bigg]
  \end{align}
where $\mathcal{I}_{k,l}(\omega)$ is a certain coefficient.
Then the sequential substitution of its right-hand side into $e^{-\frac{\pi}{2}\Lambda}$ yields
  \begin{align}
    \label{lambda-N}
    e^{-\frac{\pi}{2}\Lambda}=\sum_{k,l\geq 0}\mathcal{J}_{k,l}\qty(\frac{1}{N})^{2k+1+\frac{4l\gamma}{\pi-\gamma}},
  \end{align}
where $\mathcal{J}_{k,l}$ is a certain coefficient.

Now we can express $e_{{\rm gs},N}$ by using only $N$.
In order to obtain this expression, we use the following relations:
  \begin{align}
    \nonumber
    &\int^{\infty}_{-\infty}\rho_{\infty}(v+\Lambda)g_{+}(v)dv
    =\quad\int^{\infty}_{-\infty}\frac{g_{+}(v)}{4\cosh\frac{\pi}{2}\qty(v+\Lambda)}dv
    \\\nonumber
    &\quad=\frac{1}{2}\int_{-\infty}^{\infty}g_{+}(v)e^{-\frac{\pi}{2}\qty(v+\Lambda)}\sum_{k=0}^{\infty}\Big(-e^{-\pi(v+\Lambda)}\Big)^{k}dv
    \\
    &\quad=\frac{e^{-\frac{\pi\Lambda}{2}}}{2}\sum_{k=0}^{\infty}\tilde{g}_{+}\bigg(i\frac{\pi\qty(2k+1)}{2}\bigg)\Big(-e^{-\pi\Lambda}\Big)^{k},
  \end{align}
  \begin{align}
    \rho_{\infty}(\Lambda)
    =\frac{1}{4\cosh{\frac{\pi}{2}\Lambda}}
    =\frac{e^{-\frac{\pi}{2}\Lambda}}{2}\sum_{k=0}^{\infty}\Big(-e^{-\pi\Lambda}\Big)^{k},
  \end{align}
  \begin{align}
    \nonumber
    &\qty(\frac{1}{\rho_{N}(v)}\frac{\rm d}{{\rm d}v})^{k}\rho_{\infty}\qty(v)\Biggr|^{v=\Lambda}_{v=-\Lambda}
    \\\nonumber
&\ =\qty(\frac{1}{\rho_{N}(\Lambda)}\frac{\rm d}{{\rm d}\Lambda})^{k}\!\rho_{\infty}\qty(\Lambda)-\qty(\frac{1}{\rho_{N}(-\Lambda)}\frac{\rm d}{{\rm d}(-\Lambda)})^{k}\!\rho_{\infty}\qty(-\Lambda)
    \\
    &\ =\Big\{1-(-1)^{k}\Big\}\qty(\frac{1}{\rho_{N}(\Lambda)}\frac{\rm d}{{\rm d}\Lambda})^{k}\frac{1}{4\cosh{\frac{\pi}{2}\Lambda}}
    \\
    &\ =\Big(e^{-\frac{\pi}{2}\Lambda}\Big)^{-k}\sum_{m,l\geq 0}\mathcal{K}_{m,l}\Big(e^{-\frac{\pi}{2}\Lambda}\Big)^{2m+1+\frac{4l\gamma}{\pi-\gamma}},
  \end{align}
where $\mathcal{K}_{m,l}$ is a certain coefficient.
Therefore Eqs.~(\ref{egs_finite}) and (\ref{lambda-N}) yield
  \begin{align}
    \nonumber
    &e_{{\rm gs},N}-e_{{\rm gs},\infty}
    \\\nonumber
    &\ =2\pi A~\Biggl\{
    2\int^{\infty}_{-\infty}\rho_{\infty}(v+\Lambda)g_{+}(v)dv
    -\frac{\rho_{\infty}(\Lambda)}{N}
    \\
    &\ \quad-\frac{1}{N}\sum_{k=1}^{\infty}\frac{(-1)^{k+1}B_{k+1}}{N^{k}(k+1)}\qty(\frac{1}{\rho_{N}(v)}\frac{\rm d}{{\rm d}v})^{k}\rho_{\infty}\qty(v)\Biggr|^{v=\Lambda}_{v=-\Lambda}\Biggr\}
    \\\label{finitesize}
    &\ =\sum_{k\geq 1,m\geq 0}\mathcal{L}_{k,m}\qty(\frac{1}{N})^{2k+\frac{4m\gamma}{\pi-\gamma}}
  \end{align}
where $\mathcal{L}_{k,m}$ is a certain coefficient.
This is the finite-size corrections to the ground-state energy density of $\mathcal{H}(0)$.
By introducing the effect of $U(1)$ flux $\Phi$ into the coefficients in Eq.~(\ref{finitesize}), we get
  \begin{align}
    e_{{\rm gs},N}(\Phi)=e_{{\rm gs},\infty}+\sum_{k\geq 1,m\geq 0}\mathcal{L}_{k,m}\qty(\Phi)\qty(\frac{1}{N})^{2k+\frac{4m\gamma}{\pi-\gamma}}.
  \end{align}
Note that $e_{{\rm gs},\infty}$ is independent on $\Phi$ because this is the value in the thermodynamic limit.
Then since the inversion symmetry of the model guarantees that $e_{{\rm gs},\infty}$ is an even function of $\Phi$, we naturally expect that the difference $e_{{\rm gs},N}(\Phi)-e_{{\rm gs},N}(0)$ obeys the following finite-size scaling:
  \begin{align}
    \nonumber
    &e_{{\rm gs},N}(\Phi)-e_{{\rm gs},N}(0)
    \\\nonumber
    &\qquad=\sum_{k,l\geq1}A_{k,l}\qty(\frac{1}{N})^{2k}\Phi^{2l}
    \\\label{pre-finite}
    &\qquad\qquad\ +\sum_{k,l,m\geq1}B_{k,l,m}\qty(\frac{1}{N})^{2k+\frac{4m\gamma}{\pi-\gamma}}\Phi^{2l},
  \end{align}
where we introduced certain coefficients $A_{k,l}$ and $B_{k,l,m}$.
The above expression is intentionally split into two parts based on whether terms are including contribution from the poles of $\tilde{R}(\omega)$ dependent on the parameter $\gamma$, namely $\omega=-iq\pi\gamma/(\pi-\gamma)$ $(q\in\mathbb{N})$, or not.
Although we cannot find any constraint on the summations within this analysis, comparison of the results for the NLDWs calculated from Eq.~(\ref{pre-finite}) in the thermodynamic limit and the analytical ones~\cite{tanikawa2021exact} yields
  \begin{align}
    \nonumber
    &e_{{\rm gs},N}(\Phi)-e_{{\rm gs},N}(0)
    \\\nonumber
    &\qquad=\sum_{k\geq l\geq1}A_{k,l}\qty(\frac{1}{N})^{2k}\Phi^{2l}
    \\\label{egs-finite-corr}
    &\qquad\qquad\  +\sum_{k,l,m\geq1}B_{k,l,m}\qty(\frac{1}{N})^{2k+\frac{4m\gamma}{\pi-\gamma}}\Phi^{2l},
  \end{align}
where a new constraint on the summation appears in the first term.

\subsection*{\ref{appendixA}-3.~Case (ii): $\Delta\in$ {\Excinf}}
Here we consider the case of $\gamma=\pi r/(r+1)\ (r\in\mathbb{N})$.
Since all the values of (\ref{poles2}) are still distinct in this case, the same analysis as we have seen in case (i) is applicable.
However, there is a significant difference between case (i) and (ii), i.e., the asymptotic behavior of $R(v+2\Lambda)$.
When $\gamma=\pi r/(r+1)$, the residues of $\tilde{R}(\omega)$ at $\omega=-iq\pi\gamma/(\pi-\gamma)\ (q\in\mathbb{N})$ are
  \begin{align}
    {\rm Res}\left(\tilde{R},-i\frac{q\pi\gamma}{\pi-\gamma}\right)
    &=\frac{-i\sin\bigg(\qty(\frac{\pi}{\gamma}-2)\qty(\frac{q\pi\gamma}{\pi-\gamma})\bigg)}{2\cos{\qty(\frac{q\pi\gamma}{\pi-\gamma})}\cdot\qty(\frac{\pi}{\gamma}-1)\cos{\qty(q\pi)}}
    \\
    &=\frac{i\sin\big(\qty(r-1)q\pi\big)}{2\cos{\qty(rq\pi)}\cdot\qty(\frac{\pi}{\gamma}-1)\cos{\qty(q\pi)}}
    \\\nonumber
    &=0,
  \end{align}
which means they are no longer poles of $\tilde{R}(\omega)$.
The above fact suggests the second term in Eq.~(\ref{R(v+2lambda)}) vanishes, and thus we have
  \begin{align}
    \nonumber
    &R(v+2\Lambda)
    \\
    &=\sum_{p^{\prime}\geq1}{\rm Res}\left(\tilde{R},-i\pi\qty(p^{\prime}-\frac{1}{2})\right)\cdot\frac{e^{-\pi\qty(p^{\prime}-\frac{1}{2})(v+2\Lambda)}}{i}.
  \end{align}
for $v>0$.
Since this results in vanishing of all the terms related to the poles $\omega=-iq\pi\gamma/(\pi-\gamma)\ (q\in\mathbb{N})$, all the coefficients $B_{k,l,m}$ vanish in Eq.~(\ref{egs-finite-corr}).
Thus we naturally obtain the following finite-size scaling:
  \begin{align}
    \label{exp-special}
    e_{{\rm gs},N}(\Phi)-e_{{\rm gs},N}(0)=\sum_{k\geq l\geq1}A_{k,l}\qty(\frac{1}{N})^{2k}\Phi^{2l}.
  \end{align}
This can be understood from the perspective of the $c=1$ conformal field theory perturbed by irrelevant operators.
For examples, the above discussion is consistent with the fact that the coefficient of the umklapp term (the cosine term) vanishes at these points (see Eq.~$(2.23)$ in \cite{Lukyanov}).

\subsection*{\ref{appendixA}-4.~Case (iii): $\Delta\in$ {\Log}}
Here we consider the case of $\gamma=\pi(2p-1)/(2p-1+2q)\ (p,q\in\mathbb{N})$.
Unlike the other cases we have seen so far, some of (\ref{poles2}) take the same values, which can be written as
  \begin{align}
    \omega_{l}=-\frac{iq\pi\gamma}{\pi-\gamma}\qty(2l-1)
    =-i\pi\qty(p-\frac{1}{2})\qty(2l-1)
  \end{align}
for $l\in\mathbb{N}$.
Since this means that $\tilde{R}(\omega)$ have double poles at the above points, we have to use the following asymptotic expansion for $v>0$ instead of Eq.~(\ref{R(v+2lambda)}):
  \begin{align}
    \nonumber
    &R(v+2\Lambda)
    \\\nonumber
    &=\sideset{}{^{\prime}}\sum_{p^{\prime}\geq1}{\rm Res}\left(\tilde{R},-i\pi\qty(p^{\prime}-\frac{1}{2})\right)\cdot\frac{e^{-\pi\qty(p^{\prime}-\frac{1}{2})(v+2\Lambda)}}{i}
    \\\nonumber
    &\qquad\ +\sideset{}{^{\prime}}\sum_{q^{\prime}\geq1}{\rm Res}\left(\tilde{R},-i\frac{q^{\prime}\pi\gamma}{\pi-\gamma}\right)\cdot\frac{e^{-\frac{q^{\prime}\pi\gamma}{\pi-\gamma}(v+2\Lambda)}}{i}
    \\\label{R(v+2lambda)-beki}
    &\qquad\ +\sum_{l\geq 1}{\rm Res}\left(\tilde{R}e^{-i\omega (v+2\Lambda)},\omega_{l}\right)\cdot\frac{1}{i},
  \end{align}
where $\sum^{\prime}$ represent the summation over simple poles, namely poles excluding $\omega=\omega_{l}$.
The most different point from the other cases is the third term in Eq.~(\ref{R(v+2lambda)-beki}).
For example, we can see
  \begin{align}
    \nonumber
    &{\rm Res}\!\left(\tilde{R}e^{-i\omega (v+2\Lambda)},\omega_{1}\right)
    \\\nonumber
    &\quad={\rm Res}\!\left(\tilde{R}e^{-i\omega (v+2\Lambda)},-i\pi\qty(p-\frac{1}{2})\right)
    \\
    &\quad=\frac{-i\qty(v+2\Lambda)e^{-\pi\qty(p-\frac{1}{2})\qty(v+2\Lambda)}\sin\Big(\pi\qty(-p+\frac{1}{2}+q)\Big)}{2\sin{\Big(\pi\qty(p-\frac{1}{2})\Big)}\cdot\qty(\frac{\pi}{\gamma}-1)\cos{\qty(q\pi)}}
    \\
    &\quad=\frac{i\qty(p-\frac{1}{2})\qty(v+2\Lambda)}{2q}e^{-\pi\qty(p-\frac{1}{2})\qty(v+2\Lambda)},
  \end{align}
which reveals that a term proportional to $\Lambda e^{-\pi(2p-1)\Lambda}$ newly appears in $R\qty(v+2\Lambda)$.
This fact suggests that the previous relations (\ref{N-lambda}), (\ref{lambda-N}) and (\ref{finitesize}) are modified as follows:
  \begin{align}
    \nonumber
    \frac{1}{2N}&=\!\!\!\!\sum_{k,l,m\geq 0}\mathcal{M}_{k,l,m}\Big(e^{-\frac{\pi}{2}\Lambda}\Big)^{2k+1+\frac{4l\gamma}{\pi-\gamma}}\Big(\Lambda e^{-\pi(2p-1)\Lambda}\Big)^{m}
    \\
    &=\sum_{k,l,m\geq 0}\mathcal{M}_{k,l,m}\Big(e^{-\frac{\pi}{2}\Lambda}\Big)^{2k+1+\frac{4l\gamma}{\pi-\gamma}+2m(2p-1)}\Lambda^{m}
    \\
    \Leftrightarrow\
    &e^{-\frac{\pi}{2}\Lambda}=\!\!\!\sum_{k,l,m\geq 0}\!\!\!\mathcal{N}_{k,l,m}\qty(\frac{1}{N})^{2k+1+\frac{4l\gamma}{\pi-\gamma}+2m(2p-1)}\!\!\!\big(\log{N}\big)^{m},
    \\\nonumber
    &\!\!\!\!\!e_{{\rm gs},N}-e_{{\rm gs},\infty}
    \\
    &\ =\!\!\!\sum_{k\geq 1,~m,s\geq 0}\!\!\!\mathcal{O}_{k,m,s}\qty(\frac{1}{N})^{2k+\frac{4m\gamma}{\pi-\gamma}+2s(2p-1)}\!\!\!\big(\log{N}\big)^{s},
  \end{align}
where $\mathcal{M}_{k,l,m}$, $\mathcal{N}_{k,l,m}$ and $\mathcal{O}_{k,m,s}$ are certain coefficients.
Therefore by referring to Eq.~(\ref{egs-finite-corr}), we naturally expect that the difference $e_{{\rm gs},N}(\Phi)-e_{{\rm gs},N}(0)$ obeys the following finite-size scaling:
  \begin{align}
    \nonumber
    &e_{{\rm gs},N}(\Phi)-e_{{\rm gs},N}(0)
    \\\nonumber
    &\quad\ =\hspace{-0.4cm}\sum_{\substack{k\geq1,s\geq0\\k+s(2p-1)\geq l\geq1}}\hspace{-0.5cm}C_{k,l,s}\qty(\frac{1}{N})^{2k+2s(2p-1)}\hspace{-0.5cm}\big(\log{N}\big)^{s}\Phi^{2l}
    \\\label{finite-corr-log}
    &\qquad\quad+\hspace{-0.1cm}\sum_{\substack{k,l,m\geq1\\s\geq0}}\hspace{-0.1cm}D_{k,l,m,s}\qty(\frac{1}{N})^{2k+\frac{4m\gamma}{\pi-\gamma}+2s(2p-1)}\hspace{-0.5cm}\big(\log{N}\big)^{s}\Phi^{2l}
  \end{align}
where we introduced certain coefficients $C_{k,l,s}$ and $D_{k,l,m,s}$.

\section{The detailed analysis of the case (iii): $\Delta\in$ {\Log}}
Here we examine the NLDWs in the case (iii).
From Eq.~(\ref{finite-corr-log}), they can be calculated as
  \begin{align}
    \nonumber
    &\mathcal{D}^{(2k-1)}_N(\Theta)
    \\\nonumber
    &\quad=(2k)!\bigg[C_{k,k,0}+\chi[k-2p+1\geq1]\,C_{k-2p+1,k,1}\log{N}
    \\\label{boundary1}
    &\qquad\qquad\qquad\qquad+D_{1,k,1,0}N^{2k-2-\frac{4\gamma}{\pi-\gamma}}+\cdots\bigg],
  \end{align}
  \vspace{-0.4cm}
  \begin{align}
    \nonumber
    \mathcal{D}^{(2k)}_N(\Theta)
    &=(2k+2)!\bigg[C_{k+1,k+1,0}\frac{\Theta}{N}
    \\\nonumber
    &\qquad+\chi[k-2p+2\geq1]\,C_{k-2p+2,k+1,1}\frac{\Theta}{N}\log{N}\bigg]
    \\\label{boundary2}
    &\quad
    +Y_k(\Theta)N^{2k-1-\frac{4\gamma}{\pi-\gamma}}+\cdots,
  \end{align}
where $Y_{k}(\Theta)\equiv\sum_{l> k}(2l)!/(2l-2k-1)!D_{1,l,1,0}\Theta^{2l-2k-1}$ and $\chi[E]$ takes the value $1$ if $E$ is true and $0$ otherwise.
The greatest benefit of this analysis is that Eqs.~(\ref{boundary1}) and (\ref{boundary2}) enable us to identify the large-$N$ asymptotic behavior of the NLDWs at their boundaries between the convergent and divergent regions (see the main text).

Now we consider the behavior of the NLDWs at the points of case (iii) other than the boundary, namely $\Delta\in\big(${\Con}$\,\cup\,${\Div}$\big)\cap$ {\Log}.
Since Eqs.~(\ref{boundary1}) and (\ref{boundary2}) suggest that the effect of the logarithmic correction can appear in the NLDWs, we investigate their effect in detail.
Here we express the region with logarithmic corrections to $\mathcal{D}^{(n)}_N(\Theta)$ as $0<\gamma\leq\gamma^{(n)}_{\rm log}$, i.e., $\Delta^{(n)}_{\rm log}\leq\Delta<1$.
Since Eqs.~(\ref{boundary1}) and (\ref{boundary2}) imply that logarithmic corrections to $\mathcal{D}^{(2k-1)}_N(\Theta)$ and  $\mathcal{D}^{(2k)}_N(\Theta)$ can appear only when $1\leq p\leq\lfloor k/2\rfloor$ and $1\leq p\leq\lfloor(k+1)/2\rfloor$, respectively, the value of $\Delta^{(2k-1)}_{\rm log}$ and $\Delta^{(2k)}_{\rm log}$ can be evaluated as follows:
  \begin{align}
    &\gamma^{(2k-1)}_{\rm log}=\frac{\pi\qty(2\lfloor\frac{k}{2}\rfloor-1)}{2\lfloor\frac{k}{2}\rfloor+1}\leq\frac{\pi\qty(k-1)}{k+1}=\gamma^{(2k-1)}_{\rm B}
    \\[4pt]\label{Delta-log-2k-1}
    &\quad\Leftrightarrow\Delta^{(2k-1)}_{\rm log}\geq\Delta^{(2k-1)}_{\rm B}
  \end{align}
  \begin{align}
    &\gamma^{(2k)}_{\rm log}=\frac{\pi\qty(2\lfloor\frac{k+1}{2}\rfloor-1)}{2\lfloor\frac{k+1}{2}\rfloor+1}
    \\[4pt]
    &\qquad=
    \begin{cases}
    \frac{\pi (k-1)}{k+1}<\frac{\pi\qty(2k-1)}{2k+3}=\gamma^{(2k)}_{\rm B}& {\rm if}\ k=0~({\rm mod~} 2),
    \\[5pt]
    \frac{\pi k}{k+2}>\frac{\pi\qty(2k-1)}{2k+3}=\gamma^{(2k)}_{\rm B}& {\rm if}\ k=1~({\rm mod~} 2),
    \end{cases}
    \\\label{Delta-log-2k}
    &\quad\Leftrightarrow
    \begin{cases}
    \Delta^{(2k)}_{\rm log}>\Delta^{(2k)}_{\rm B}& {\rm if}\ k=0~({\rm mod~} 2),
    \\[5pt]
    \Delta^{(2k)}_{\rm log}<\Delta^{(2k)}_{\rm B}& {\rm if}\ k=1~({\rm mod~} 2),
    \end{cases}
  \end{align}
where $\gamma^{(n)}_{\rm B}=\arccos{\Delta^{(n)}_{\rm B}}$ and $\lfloor x\rfloor$ is the floor function.
The equal sign in Eq.~(\ref{Delta-log-2k-1}) holds when $k$ is even.
Note that $\gamma^{(2k-1)}_{\rm log}$ and $\gamma^{(2k)}_{\rm log}$ correspond to the cases when $p=\lfloor k/2\rfloor,\,q=1$ and $p=\lfloor (k+1)/2\rfloor,\,q=1$, respectively, as $\gamma=\pi(2p-1)/(2p-1+2q)$ is monotonically increasing for $p$ and decreasing for $q$.

Based on the above results, we can specify the large-$N$ asymptotic behaviours of $\mathcal{D}^{(n)}_N(\Theta)$ at $\Delta\in\big(${\Con}$\,\cup\,${\Div}$\big)\cap$ {\Log}.
From Eq.~(\ref{Delta-log-2k-1}), we can see that logarithmic corrections to $\mathcal{D}^{(2k-1)}_N(\Theta)$ can appear only in the divergent region {$\mathcal{S}_{d}^{(2k-1)}$}.
Nevertheless, the leading term of $\mathcal{D}^{(2k-1)}_N(\Theta)$ in this region is $N^{2k-2-4\gamma/(\pi-\gamma)}$ as can be seen from Eq.~(\ref{boundary1}).
Thus in the large-$N$ limit $\mathcal{D}^{(2k-1)}_N(\Theta)$ shows the power-law divergence after all.
In the convergent region {$\mathcal{S}_{c}^{(2k-1)}$}, the absence of the logarithmic corrections reproduces the same behaviour as Eq.~(\ref{D2n-1}), and thus $\mathcal{D}^{(2k-1)}_N(\Theta)$ shows the convergence to a finite value in the thermodynamic limit (see Table~\ref{table-group-all}).
Similarly, although there exist the logarithmic corrections, $\mathcal{D}^{(2k)}_N(\Theta)$ shows, in the large-$N$ limit, the power-law divergence of the form $\mathcal{D}^{(2k)}_N(\Theta)\sim N^{2k-1-4\gamma/(\pi-\gamma)}$ in the divergent region {$\mathcal{S}_{d}^{(2k)}$} as well.
In the convergent region {$\mathcal{S}_{c}^{(2k)}$}, when $k=0$ $($mod $2)$, the absence of the logarithmic corrections reproduces the same behaviour as Eq.~(\ref{D2n}), and thus $\mathcal{D}^{(2k)}_N(\Theta)$ vanishes in the thermodynamic limit.
On the other hand, when $k=1$ $($mod $2)$, Eq.~(\ref{Delta-log-2k}) suggests that the logarithmic corrections to $\mathcal{D}^{(2k)}_N(\Theta)$ can appear even in the convergent region {$\mathcal{S}_{c}^{(2k)}$}, namely in $\Delta_{\rm log}^{(2k)}\leq\Delta<\Delta^{(2k)}_{\rm B}$ \big(see Eq.~{(\ref{D21})} for example\big).
However, since the leading order of $\mathcal{D}^{(2k)}_N(\Theta)$ can be written as $\mathcal{O}\qty(N^{-1}\log{N},1/N^{-2k+1+\frac{4\gamma}{\pi-\gamma}})$ in this case, $\mathcal{D}^{(2k)}_N(\Theta)$ still vanishes in the thermodynamic limit as we have seen in the other case (see Table~\ref{table-group-all}).

\section{Numerical results for the finite-size corrections}
\label{appendixB}
We show some numerical results for the finite-size corrections with the $U(1)$ flux.
Since the expansion (\ref{pre-finite}) was studied in our previous paper~\cite{tanikawa2021exact}, here we focus only on the new result (\ref{finite-corr-log}).
In order to confirm the finite-size scaling of $e_{{\rm gs},N}(\Phi)$, it is better to calculate the NLDWs instead of  $e_{{\rm gs},N}(\Phi)$ itself.
Below we discuss several dominant terms of the NLDWs in certain cases.

\begin{figure*}[htbp]
    \includegraphics[width=\hsize]{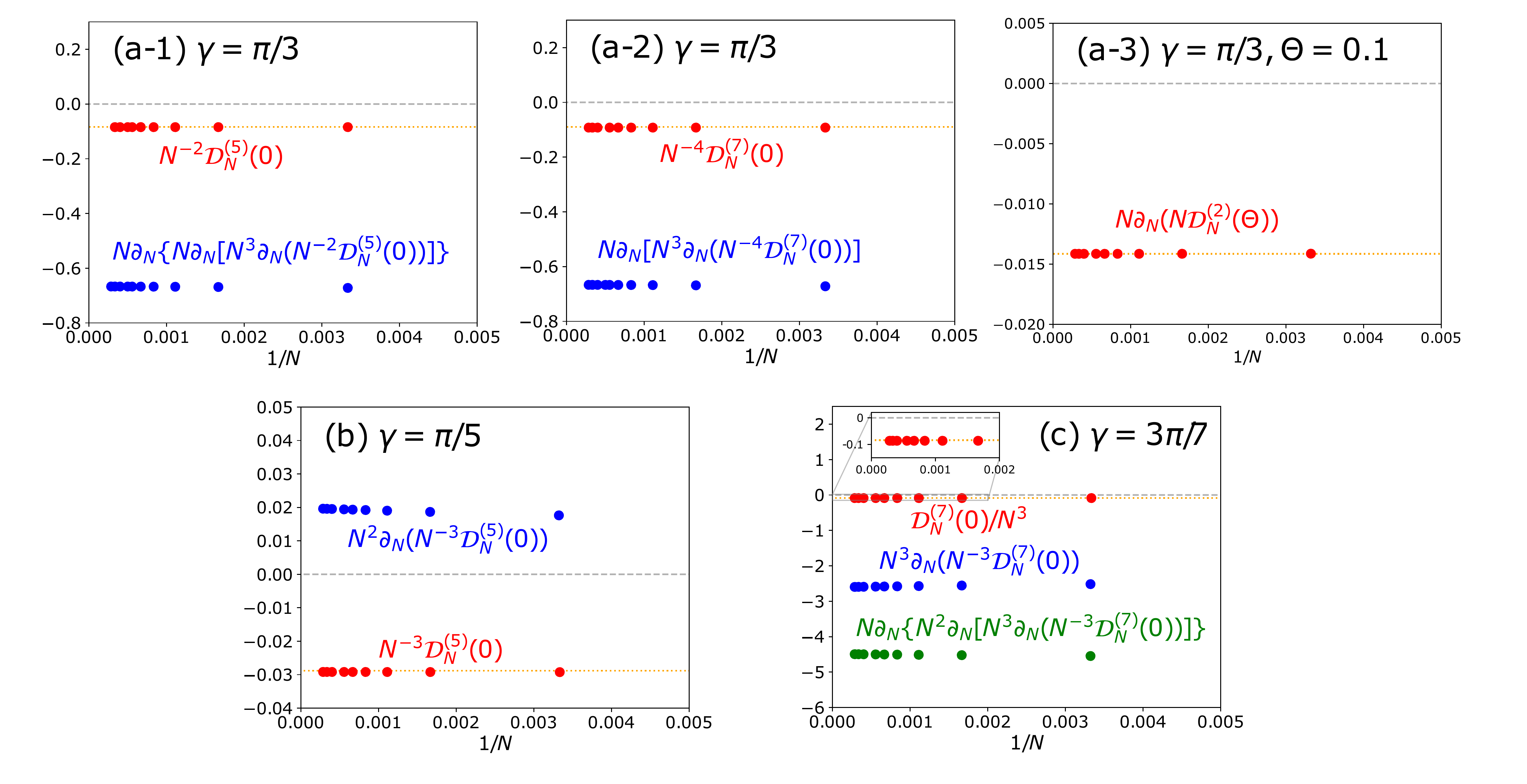}
    \vspace{-0.6cm}
    \caption{Numerical results for $N$ dependence of the quantities related to the NLDWs.
    They have been studied for system sizes ranging from $N=300$ up to $N=3500$.
    All the vertical axes are scaled with $J$.
    Note that the data shown in red are related to the leading terms and those in blue (green) related to the sub- (sub-sub-)leading terms.
    We can see that the data for each quantity fall on an almost straight line to a finite value in the large-$N$ region.
    The extrapolated values are (a-1) [red] $-0.08451...$ and [blue] $-0.6446...$, (a-2) [red] $-0.09226...$ and [blue] $-0.6666...$, (a-3) [red] $-0.01414...$, (b) [red] $-0.02918...$ and [blue] $0.01983...$, and (c) [red] $-0.08650...$, [blue] $-2.604...$ and [green] $-4.488...$.
    The orange dotted lines indicate the analytical values in the thermodynamic limit.}
    \label{D51-D72}
\end{figure*}

\subsection*{\ref{appendixB}-1.~At $\gamma=\frac{\pi}{3}$ $(p=q=1)$}
We show the results for three examples:~the second-order, fifth-order and seventh-order ones.

First, we consider the fifth-order and seventh-order ones, which diverge in the thermodynamic limit at this point.
From Eq.~(\ref{boundary1}), they can be calculated as
  \begin{align}
    \label{D51}
    \mathcal{D}^{(5)}_N(0)&=6!D_{1,3,1,0}N^2+6!C_{1,3,2}\big(\log{N}\big)^2+\mathcal{O}\big(\log{N}\big)
    \\\label{D71}
    \mathcal{D}^{(7)}_N(0)&\!=\!8!D_{1,4,1,0}N^4\!+\!8!D_{1,4,1,1}N^2\log{N}\!+\!\mathcal{O}\qty(\big(\log{N}\big)^3).
  \end{align}
Since both of their leading behaviors are the power-law divergence, the coefficients can be estimated through
  \begin{align}
    N^{-2}\mathcal{D}^{(5)}_N(0)
    &=6!D_{1,3,1,0}+\mathcal{O}\left(\qty(\frac{\log{N}}{N})^2\right)
    \\
    N^{-4}\mathcal{D}^{(7)}_N(0)
    &=8!D_{1,4,1,0}+\mathcal{O}\left(\frac{\log{N}}{N^2}\right).
  \end{align}
On the other hand, in order to obtain their sub-leading behaviors, it is useful to differentiate them with respect to $N$.
Therefore we introduce the following quantities:
  \begin{align}
    \nonumber
    N\partial_{N}&\Bigg\{N\partial_{N}\bigg[N^3\partial_{N}\Big(N^{-2}\mathcal{D}^{(5)}_N(0)\Big)\bigg]\Bigg\}
    \\
    &\qquad=-4\cdot 6!C_{1,3,2}+\mathcal{O}\left(\frac{\big(\log{N}\big)^{3}}{N^2}\right)
    \\\nonumber
    N\partial_{N}&\bigg[N^3\partial_{N}\Big(N^{-4}\mathcal{D}^{(7)}_N(0)\Big)\bigg]
    \\
    &\qquad=-2\cdot 8!D_{1,4,1,1}+\mathcal{O}\left(\frac{\big(\log{N}\big)^{3}}{N^2}\right).
  \end{align}
The numerical results for the above quantities are shown in Figs.~\ref{D51-D72}(a-1) and (a-2).
Note that, from Eq.~(4.1) in Ref.~\cite{Lukyanov}, the analytical values at $\gamma=\pi/3$ can be calculated as $6!D_{1,3,1,0}/J=-0.08360...$ and $8!D_{1,4,1,0}/J=-0.08968...$ which  are indicated by the orange dotted lines.
These figures obviously show that the data fall on straight lines to finite values in the large-$N$ region, which confirms Eqs.~(\ref{D51}) and (\ref{D71}).

Finally, we consider the second-order one, which converge in the thermodynamic limit at this point.
From Eq.~(\ref{boundary2}), this can be calculated as
  \begin{align}
    \label{D21}
    \mathcal{D}^{(2)}_N(\Theta)&=4!C_{1,2,1}\frac{\Theta}{N}\log{N}+\mathcal{O}\qty(\frac{1}{N})
    \\
    &=-\frac{81\sqrt{3}J}{32\pi^3}\frac{\Theta}{N}\log{N}+\mathcal{O}\qty(\frac{1}{N})
  \end{align}
where we obtained $C_{1,2,1}$ at $\gamma=\pi/3$ from Eq.~(4.1) in Ref.~\cite{Lukyanov}.
Then, in order to evaluate the leading behavior, we introduce the following quantity:
  \begin{align}
    N\partial_{N}\left(N \mathcal{D}^{(2)}_N(\Theta)\right)=-\frac{81\sqrt{3}J}{32\pi^3}\Theta+\mathcal{O}\qty(\bigg(\frac{\log{N}}{N}\bigg)^2).
  \end{align}
The numerical results for the above quantity is shown in Fig.\ref{D51-D72}(a-3).
This figure clearly shows that, in the large-$N$ region, the data fall on a straight line to the analytical value $-81\sqrt{3}/(32\pi^3)\Theta=-0.01414...$ indicated by the orange dotted line, which confirms Eq.~(\ref{D21}).

\subsection*{\ref{appendixB}-2.~At $\gamma=\frac{\pi}{5}$ $(p=1,q=2)$}
We consider the fifth-order one.
From Eq.~(\ref{boundary1}), this can be calculated as
  \begin{align}
    \label{D52}
    \mathcal{D}^{(5)}_N(0)&=6!D_{1,3,1,0}N^3+6!D_{1,3,2,0}N^2+\mathcal{O}\big(N\log{N}\big).
  \end{align}
As in the previous case, we introduce the following quantities:
  \begin{align}
    N^{-3}\mathcal{D}^{(5)}_N(0)
    &=6!D_{1,3,1,0}+\mathcal{O}\qty(\frac{1}{N})
    \\
    N^2\partial_{N}\Big(N^{-3}\mathcal{D}^{(5)}_N(0)\Big)&=-6!D_{1,3,2,0}+\mathcal{O}\left(\frac{\log{N}}{N}\right).
  \end{align}
The numerical results for the above quantities are shown in Fig.~\ref{D51-D72}(b).
Since the analytical form of $D_{1,k,1,0}$ can be obtained as
\begin{align}
  \nonumber
  &D_{1,k,1,0}=
  -\frac{16\pi J\sin{\gamma}\sin{\big(\frac{2\pi\gamma}{\pi-\gamma}\big)}}{(2k)!\,\gamma}
  \cdot
  \frac{\Gamma^2{\big(\frac{\pi}{\pi-\gamma}\big)}\Gamma^2{\big(-\frac{2\pi}{\pi-\gamma}\big)}}
  {\Gamma^2{\big(-\frac{\pi}{\pi-\gamma}\big)}}
  \\\label{D1k1}
  &\quad\quad
  \Bigg[
  \frac{(\pi-\gamma)\,\Gamma{\big(\frac{\pi-\gamma}{2\gamma}\big)}}{\sqrt{\pi}\,\Gamma{\big(\frac{\pi}{2\gamma}\big)}}
  \Bigg]^{\frac{4\gamma}{\pi-\gamma}}
  \!\!
  \dv[2k]{\Phi}
  \frac{\Gamma{\big(\frac{\Phi+2\pi}{2(\pi-\gamma)}\big)}\Gamma{\big(\frac{-\Phi+2\pi}{2(\pi-\gamma)}\big)}}{\Gamma{\big(\frac{\Phi-2\gamma}{2(\pi-\gamma)}\big)}\Gamma{\big(\frac{-\Phi-2\gamma}{2(\pi-\gamma)}\big)}}\Biggr|_{\Phi=0}
\end{align}
from Eq.~(4.1) in Ref.~\cite{Lukyanov}, we have $6!D_{1,3,1,0}/J=-0.02881...$ at $\gamma=\pi/5$ which is indicated by the orange dotted line.
This figure obviously shows that the data fall on straight lines to finite values in the large-$N$ region, which confirms Eq.~(\ref{D52}).

\subsection*{\ref{appendixB}-3.~At $\gamma=\frac{3\pi}{7}$ $(p=2,q=2)$}
We consider the seventh-order one.
Here we focus not only on the leading and sub-leading term but also on the sub-sub-leading term.
From Eq.~(\ref{boundary1}), this can be calculated as
  \begin{align}
    \nonumber
    \mathcal{D}^{(7)}_N(0)
    =8!&D_{1,4,1,0}N^3+8!D_{2,4,1,0}N
    \\\label{D72}
    &\qquad+8!C_{1,4,1}\log{N}+\mathcal{O}\qty(\frac{1}{N}).
  \end{align}
As in the previous case, we introduce the following quantities:
  \begin{align}
    N^{-3}\mathcal{D}^{(7)}_N(0)
    &=8!D_{1,4,1,0}+\mathcal{O}\qty(\frac{1}{N^2})
    \\
    N^3\partial_{N}\Big(N^{-3}\mathcal{D}^{(7)}_N(0)\Big)
    &=-2\cdot 8!D_{2,4,1,0}+\mathcal{O}\qty(\frac{\log{N}}{N})
    \\\nonumber
    N\partial_{N}\Bigg\{N^2\partial_{N}\bigg[N^{3}\partial_{N}&\Big(N^{-3}\mathcal{D}^{(7)}_N(0)\Big)\bigg]\Bigg\}
    \\
    &=3\cdot 8!C_{1,4,1}+\mathcal{O}\qty(\frac{1}{N}).
  \end{align}
The numerical results for the above quantities are shown in Fig.~\ref{D51-D72}(c).
From Eq.~(\ref{D1k1}), we have $8!D_{1,4,1,0}/J=-0.08433...$ at $\gamma=3\pi/7$ which is indicated by the orange dotted line.
The figure obviously shows that the data fall on straight lines to finite values in the large-$N$ region, which confirms Eq.~(\ref{D72}).

\bibliographystyle{apsrev4-1}

\begin{thebibliography}{47}%
\makeatletter
\providecommand \@ifxundefined [1]{%
 \@ifx{#1\undefined}
}%
\providecommand \@ifnum [1]{%
 \ifnum #1\expandafter \@firstoftwo
 \else \expandafter \@secondoftwo
 \fi
}%
\providecommand \@ifx [1]{%
 \ifx #1\expandafter \@firstoftwo
 \else \expandafter \@secondoftwo
 \fi
}%
\providecommand \natexlab [1]{#1}%
\providecommand \enquote  [1]{``#1''}%
\providecommand \bibnamefont  [1]{#1}%
\providecommand \bibfnamefont [1]{#1}%
\providecommand \citenamefont [1]{#1}%
\providecommand \href@noop [0]{\@secondoftwo}%
\providecommand \href [0]{\begingroup \@sanitize@url \@href}%
\providecommand \@href[1]{\@@startlink{#1}\@@href}%
\providecommand \@@href[1]{\endgroup#1\@@endlink}%
\providecommand \@sanitize@url [0]{\catcode `\\12\catcode `\$12\catcode
  `\&12\catcode `\#12\catcode `\^12\catcode `\_12\catcode `\%12\relax}%
\providecommand \@@startlink[1]{}%
\providecommand \@@endlink[0]{}%
\providecommand \url  [0]{\begingroup\@sanitize@url \@url }%
\providecommand \@url [1]{\endgroup\@href {#1}{\urlprefix }}%
\providecommand \urlprefix  [0]{URL }%
\providecommand \Eprint [0]{\href }%
\providecommand \doibase [0]{http://dx.doi.org/}%
\providecommand \selectlanguage [0]{\@gobble}%
\providecommand \bibinfo  [0]{\@secondoftwo}%
\providecommand \bibfield  [0]{\@secondoftwo}%
\providecommand \translation [1]{[#1]}%
\providecommand \BibitemOpen [0]{}%
\providecommand \bibitemStop [0]{}%
\providecommand \bibitemNoStop [0]{.\EOS\space}%
\providecommand \EOS [0]{\spacefactor3000\relax}%
\providecommand \BibitemShut  [1]{\csname bibitem#1\endcsname}%
\let\auto@bib@innerbib\@empty
\bibitem [{\citenamefont {Zotos}(2005)}]{zotos2005}%
  \BibitemOpen
  \bibfield  {author} {\bibinfo {author} {\bibfnamefont {X.}~\bibnamefont
  {Zotos}},\ }\href {\doibase 10.1143/JPSJS.74S.173} {\bibfield  {journal}
  {\bibinfo  {journal} {J. Phys. Soc. Jpn.}\ }\textbf {\bibinfo {volume}
  {74}},\ \bibinfo {pages} {173} (\bibinfo {year} {2005})}\BibitemShut
  {NoStop}%
\bibitem [{\citenamefont {Zotos}\ and\ \citenamefont
  {Prelov{\v{s}}ek}(2004)}]{zotos2004transport}%
  \BibitemOpen
  \bibfield  {author} {\bibinfo {author} {\bibfnamefont {X.}~\bibnamefont
  {Zotos}}\ and\ \bibinfo {author} {\bibfnamefont {P.}~\bibnamefont
  {Prelov{\v{s}}ek}},\ }\href@noop {} {\emph {\bibinfo {title} {Transport in
  one dimensional quantum systems, {\rm in} Strong interactions in low
  dimensions}}}\ (\bibinfo  {publisher} {Kluwer Academic Publishers,
  Doodrecht},\ \bibinfo {year} {2004})\BibitemShut {NoStop}%
\bibitem [{\citenamefont {Bertini}\ \emph {et~al.}(2021)\citenamefont
  {Bertini}, \citenamefont {Heidrich-Meisner}, \citenamefont {Karrasch},
  \citenamefont {Prosen}, \citenamefont {Steinigeweg},\ and\ \citenamefont
  {\ifmmode \check{Z}\else \v{Z}\fi{}nidari\ifmmode~\check{c}\else
  \v{c}\fi{}}}]{Bertini2020}%
  \BibitemOpen
  \bibfield  {author} {\bibinfo {author} {\bibfnamefont {B.}~\bibnamefont
  {Bertini}}, \bibinfo {author} {\bibfnamefont {F.}~\bibnamefont
  {Heidrich-Meisner}}, \bibinfo {author} {\bibfnamefont {C.}~\bibnamefont
  {Karrasch}}, \bibinfo {author} {\bibfnamefont {T.}~\bibnamefont {Prosen}},
  \bibinfo {author} {\bibfnamefont {R.}~\bibnamefont {Steinigeweg}}, \ and\
  \bibinfo {author} {\bibfnamefont {M.}~\bibnamefont {\ifmmode \check{Z}\else
  \v{Z}\fi{}nidari\ifmmode~\check{c}\else \v{c}\fi{}}},\ }\href {\doibase
  10.1103/RevModPhys.93.025003} {\bibfield  {journal} {\bibinfo  {journal}
  {Rev. Mod. Phys.}\ }\textbf {\bibinfo {volume} {93}},\ \bibinfo {pages}
  {025003} (\bibinfo {year} {2021})}\BibitemShut {NoStop}%
\bibitem [{\citenamefont {Sirker}(2020)}]{Sirker2020}%
  \BibitemOpen
  \bibfield  {author} {\bibinfo {author} {\bibfnamefont {J.}~\bibnamefont
  {Sirker}},\ }\href {\doibase 10.21468/SciPostPhysLectNotes.17} {\bibfield
  {journal} {\bibinfo  {journal} {SciPost Phys. Lect. Notes}\ ,\ \bibinfo
  {pages} {17}} (\bibinfo {year} {2020})}\BibitemShut {NoStop}%
\bibitem [{\citenamefont {Hirobe}\ \emph {et~al.}(2017)\citenamefont {Hirobe},
  \citenamefont {Sato}, \citenamefont {Kawamata} \emph
  {et~al.}}]{hirobe2017one}%
  \BibitemOpen
  \bibfield  {author} {\bibinfo {author} {\bibfnamefont {D.}~\bibnamefont
  {Hirobe}}, \bibinfo {author} {\bibfnamefont {M.}~\bibnamefont {Sato}},
  \bibinfo {author} {\bibfnamefont {T.}~\bibnamefont {Kawamata}},  \emph
  {et~al.},\ }\href@noop {} {\bibfield  {journal} {\bibinfo  {journal} {Nature
  Phys}\ }\textbf {\bibinfo {volume} {13}},\ \bibinfo {pages} {30} (\bibinfo
  {year} {2017})}\BibitemShut {NoStop}%
\bibitem [{\citenamefont {Kubo}(1957)}]{Kubo}%
  \BibitemOpen
  \bibfield  {author} {\bibinfo {author} {\bibfnamefont {R.}~\bibnamefont
  {Kubo}},\ }\href@noop {} {\bibfield  {journal} {\bibinfo  {journal} {J. Phys.
  Soc. Jpn.}\ }\textbf {\bibinfo {volume} {12}},\ \bibinfo {pages} {570}
  (\bibinfo {year} {1957})}\BibitemShut {NoStop}%
\bibitem [{\citenamefont {Watanabe}\ and\ \citenamefont
  {Oshikawa}(2020)}]{Watanabe-Oshikawa}%
  \BibitemOpen
  \bibfield  {author} {\bibinfo {author} {\bibfnamefont {H.}~\bibnamefont
  {Watanabe}}\ and\ \bibinfo {author} {\bibfnamefont {M.}~\bibnamefont
  {Oshikawa}},\ }\href {\doibase 10.1103/PhysRevB.102.165137} {\bibfield
  {journal} {\bibinfo  {journal} {Phys. Rev. B}\ }\textbf {\bibinfo {volume}
  {102}},\ \bibinfo {pages} {165137} (\bibinfo {year} {2020})}\BibitemShut
  {NoStop}%
\bibitem [{\citenamefont {Watanabe}\ \emph {et~al.}(2020)\citenamefont
  {Watanabe}, \citenamefont {Liu},\ and\ \citenamefont
  {Oshikawa}}]{Watanabe-Oshikawa-Liu}%
  \BibitemOpen
  \bibfield  {author} {\bibinfo {author} {\bibfnamefont {H.}~\bibnamefont
  {Watanabe}}, \bibinfo {author} {\bibfnamefont {Y.}~\bibnamefont {Liu}}, \
  and\ \bibinfo {author} {\bibfnamefont {M.}~\bibnamefont {Oshikawa}},\ }\href
  {\doibase 10.1007/s10955-020-02654-5} {\bibfield  {journal} {\bibinfo
  {journal} {J. Stat. Phys.}\ }\textbf {\bibinfo {volume} {181}},\ \bibinfo
  {pages} {2050} (\bibinfo {year} {2020})}\BibitemShut {NoStop}%
\bibitem [{\citenamefont {Kohn}(1964)}]{Kohn}%
  \BibitemOpen
  \bibfield  {author} {\bibinfo {author} {\bibfnamefont {W.}~\bibnamefont
  {Kohn}},\ }\href@noop {} {\bibfield  {journal} {\bibinfo  {journal} {Phys.
  Rev.}\ }\textbf {\bibinfo {volume} {133}},\ \bibinfo {pages} {A171} (\bibinfo
  {year} {1964})}\BibitemShut {NoStop}%
\bibitem [{\citenamefont {Fye}\ \emph {et~al.}(1991)\citenamefont {Fye},
  \citenamefont {Martins}, \citenamefont {Scalapino}, \citenamefont {Wagner},\
  and\ \citenamefont {Hanke}}]{Fye1991}%
  \BibitemOpen
  \bibfield  {author} {\bibinfo {author} {\bibfnamefont {R.~M.}\ \bibnamefont
  {Fye}}, \bibinfo {author} {\bibfnamefont {M.~J.}\ \bibnamefont {Martins}},
  \bibinfo {author} {\bibfnamefont {D.~J.}\ \bibnamefont {Scalapino}}, \bibinfo
  {author} {\bibfnamefont {J.}~\bibnamefont {Wagner}}, \ and\ \bibinfo {author}
  {\bibfnamefont {W.}~\bibnamefont {Hanke}},\ }\href {\doibase
  10.1103/PhysRevB.44.6909} {\bibfield  {journal} {\bibinfo  {journal} {Phys.
  Rev. B}\ }\textbf {\bibinfo {volume} {44}},\ \bibinfo {pages} {6909}
  (\bibinfo {year} {1991})}\BibitemShut {NoStop}%
\bibitem [{\citenamefont {Stafford}\ and\ \citenamefont
  {Millis}(1993)}]{Stafford1993}%
  \BibitemOpen
  \bibfield  {author} {\bibinfo {author} {\bibfnamefont {C.~A.}\ \bibnamefont
  {Stafford}}\ and\ \bibinfo {author} {\bibfnamefont {A.~J.}\ \bibnamefont
  {Millis}},\ }\href {\doibase 10.1103/PhysRevB.48.1409} {\bibfield  {journal}
  {\bibinfo  {journal} {Phys. Rev. B}\ }\textbf {\bibinfo {volume} {48}},\
  \bibinfo {pages} {1409} (\bibinfo {year} {1993})}\BibitemShut {NoStop}%
\bibitem [{\citenamefont {Fujimoto}\ and\ \citenamefont
  {Kawakami}(2003)}]{fujimoto2003}%
  \BibitemOpen
  \bibfield  {author} {\bibinfo {author} {\bibfnamefont {S.}~\bibnamefont
  {Fujimoto}}\ and\ \bibinfo {author} {\bibfnamefont {N.}~\bibnamefont
  {Kawakami}},\ }\href {\doibase 10.1103/PhysRevLett.90.197202} {\bibfield
  {journal} {\bibinfo  {journal} {Phys. Rev. Lett.}\ }\textbf {\bibinfo
  {volume} {90}},\ \bibinfo {pages} {197202} (\bibinfo {year}
  {2003})}\BibitemShut {NoStop}%
\bibitem [{\citenamefont {Kirchner}\ \emph {et~al.}(1999)\citenamefont
  {Kirchner}, \citenamefont {Evertz},\ and\ \citenamefont
  {Hanke}}]{Kirchner1999}%
  \BibitemOpen
  \bibfield  {author} {\bibinfo {author} {\bibfnamefont {S.}~\bibnamefont
  {Kirchner}}, \bibinfo {author} {\bibfnamefont {H.~G.}\ \bibnamefont
  {Evertz}}, \ and\ \bibinfo {author} {\bibfnamefont {W.}~\bibnamefont
  {Hanke}},\ }\href {\doibase 10.1103/PhysRevB.59.1825} {\bibfield  {journal}
  {\bibinfo  {journal} {Phys. Rev. B}\ }\textbf {\bibinfo {volume} {59}},\
  \bibinfo {pages} {1825} (\bibinfo {year} {1999})}\BibitemShut {NoStop}%
\bibitem [{\citenamefont {Sirker}\ \emph {et~al.}(2009)\citenamefont {Sirker},
  \citenamefont {Pereira},\ and\ \citenamefont {Affleck}}]{Sirker2009}%
  \BibitemOpen
  \bibfield  {author} {\bibinfo {author} {\bibfnamefont {J.}~\bibnamefont
  {Sirker}}, \bibinfo {author} {\bibfnamefont {R.~G.}\ \bibnamefont {Pereira}},
  \ and\ \bibinfo {author} {\bibfnamefont {I.}~\bibnamefont {Affleck}},\ }\href
  {\doibase 10.1103/PhysRevLett.103.216602} {\bibfield  {journal} {\bibinfo
  {journal} {Phys. Rev. Lett.}\ }\textbf {\bibinfo {volume} {103}},\ \bibinfo
  {pages} {216602} (\bibinfo {year} {2009})}\BibitemShut {NoStop}%
\bibitem [{\citenamefont {Urichuk}\ \emph {et~al.}(2021)\citenamefont
  {Urichuk}, \citenamefont {Sirker},\ and\ \citenamefont
  {Kl\"umper}}]{Urichuk2021}%
  \BibitemOpen
  \bibfield  {author} {\bibinfo {author} {\bibfnamefont {A.}~\bibnamefont
  {Urichuk}}, \bibinfo {author} {\bibfnamefont {J.}~\bibnamefont {Sirker}}, \
  and\ \bibinfo {author} {\bibfnamefont {A.}~\bibnamefont {Kl\"umper}},\ }\href
  {\doibase 10.1103/PhysRevB.103.245108} {\bibfield  {journal} {\bibinfo
  {journal} {Phys. Rev. B}\ }\textbf {\bibinfo {volume} {103}},\ \bibinfo
  {pages} {245108} (\bibinfo {year} {2021})}\BibitemShut {NoStop}%
\bibitem [{\citenamefont {Tanikawa}\ \emph {et~al.}(2021)\citenamefont
  {Tanikawa}, \citenamefont {Takasan},\ and\ \citenamefont
  {Katsura}}]{tanikawa2021exact}%
  \BibitemOpen
  \bibfield  {author} {\bibinfo {author} {\bibfnamefont {Y.}~\bibnamefont
  {Tanikawa}}, \bibinfo {author} {\bibfnamefont {K.}~\bibnamefont {Takasan}}, \
  and\ \bibinfo {author} {\bibfnamefont {H.}~\bibnamefont {Katsura}},\ }\href
  {\doibase 10.1103/PhysRevB.103.L201120} {\bibfield  {journal} {\bibinfo
  {journal} {Phys. Rev. B}\ }\textbf {\bibinfo {volume} {103}},\ \bibinfo
  {pages} {L201120} (\bibinfo {year} {2021})}\BibitemShut {NoStop}%
\bibitem [{\citenamefont {Fava}\ \emph {et~al.}(2021)\citenamefont {Fava},
  \citenamefont {Biswas}, \citenamefont {Gopalakrishnan}, \citenamefont
  {Vasseur},\ and\ \citenamefont {Parameswaran}}]{fava2021hydrodynamic}%
  \BibitemOpen
  \bibfield  {author} {\bibinfo {author} {\bibfnamefont {M.}~\bibnamefont
  {Fava}}, \bibinfo {author} {\bibfnamefont {S.}~\bibnamefont {Biswas}},
  \bibinfo {author} {\bibfnamefont {S.}~\bibnamefont {Gopalakrishnan}},
  \bibinfo {author} {\bibfnamefont {R.}~\bibnamefont {Vasseur}}, \ and\
  \bibinfo {author} {\bibfnamefont {S.~A.}\ \bibnamefont {Parameswaran}},\
  }\href@noop {} {\enquote {\bibinfo {title} {Hydrodynamic non-linear response
  of interacting integrable systems},}\ } (\bibinfo {year} {2021}),\ \Eprint
  {http://arxiv.org/abs/2103.06899} {arXiv:2103.06899 [cond-mat.str-el]}
  \BibitemShut {NoStop}%
\bibitem [{\citenamefont {Takasan}\ \emph {et~al.}(2021)\citenamefont
  {Takasan}, \citenamefont {Oshikawa},\ and\ \citenamefont
  {Watanabe}}]{takasan2021adiabatic}%
  \BibitemOpen
  \bibfield  {author} {\bibinfo {author} {\bibfnamefont {K.}~\bibnamefont
  {Takasan}}, \bibinfo {author} {\bibfnamefont {M.}~\bibnamefont {Oshikawa}}, \
  and\ \bibinfo {author} {\bibfnamefont {H.}~\bibnamefont {Watanabe}},\
  }\href@noop {} {\enquote {\bibinfo {title} {Adiabatic transport in
  one-dimensional systems with a single defect},}\ } (\bibinfo {year} {2021}),\
  \Eprint {http://arxiv.org/abs/2105.11378} {arXiv:2105.11378
  [cond-mat.mes-hall]} \BibitemShut {NoStop}%
\bibitem [{\citenamefont {Fukusumi}\ and\ \citenamefont
  {Barišić}(2021)}]{fukusumi2021kubos}%
  \BibitemOpen
  \bibfield  {author} {\bibinfo {author} {\bibfnamefont {Y.}~\bibnamefont
  {Fukusumi}}\ and\ \bibinfo {author} {\bibfnamefont {O.~S.}\ \bibnamefont
  {Barišić}},\ }\href@noop {} {\enquote {\bibinfo {title} {Kubo's response
  theory and bosonization with a background gauge field and irrelevant
  perturbations},}\ } (\bibinfo {year} {2021}),\ \Eprint
  {http://arxiv.org/abs/2106.07339} {arXiv:2106.07339 [cond-mat.stat-mech]}
  \BibitemShut {NoStop}%
\bibitem [{\citenamefont {Takahashi}(2005)}]{Takahashi}%
  \BibitemOpen
  \bibfield  {author} {\bibinfo {author} {\bibfnamefont {M.}~\bibnamefont
  {Takahashi}},\ }\href@noop {} {\emph {\bibinfo {title} {Thermodynamics of
  one-dimensional solvable models}}}\ (\bibinfo  {publisher} {Cambridge
  university press},\ \bibinfo {year} {2005})\BibitemShut {NoStop}%
\bibitem [{\citenamefont {Korepin}\ \emph {et~al.}(1993)\citenamefont
  {Korepin}, \citenamefont {Bogoliubov},\ and\ \citenamefont
  {Izergin}}]{Korepin_book}%
  \BibitemOpen
  \bibfield  {author} {\bibinfo {author} {\bibfnamefont {V.~E.}\ \bibnamefont
  {Korepin}}, \bibinfo {author} {\bibfnamefont {N.~M.}\ \bibnamefont
  {Bogoliubov}}, \ and\ \bibinfo {author} {\bibfnamefont {A.~G.}\ \bibnamefont
  {Izergin}},\ }\href {\doibase 10.1017/CBO9780511628832} {\emph {\bibinfo
  {title} {Quantum Inverse Scattering Method and Correlation Functions}}},\
  Cambridge Monographs on Mathematical Physics\ (\bibinfo  {publisher}
  {Cambridge University Press},\ \bibinfo {year} {1993})\BibitemShut {NoStop}%
\bibitem [{\citenamefont {Yang}\ and\ \citenamefont
  {Yang}(1966{\natexlab{a}})}]{yang1966two}%
  \BibitemOpen
  \bibfield  {author} {\bibinfo {author} {\bibfnamefont {C.~N.}\ \bibnamefont
  {Yang}}\ and\ \bibinfo {author} {\bibfnamefont {C.~P.}\ \bibnamefont
  {Yang}},\ }\href@noop {} {\bibfield  {journal} {\bibinfo  {journal} {Phys.
  Rev.}\ }\textbf {\bibinfo {volume} {150}},\ \bibinfo {pages} {327} (\bibinfo
  {year} {1966}{\natexlab{a}})}\BibitemShut {NoStop}%
\bibitem [{\citenamefont {Hamer}\ \emph {et~al.}(1987)\citenamefont {Hamer},
  \citenamefont {Quispel},\ and\ \citenamefont {Batchelor}}]{Hamer}%
  \BibitemOpen
  \bibfield  {author} {\bibinfo {author} {\bibfnamefont {C.}~\bibnamefont
  {Hamer}}, \bibinfo {author} {\bibfnamefont {G.}~\bibnamefont {Quispel}}, \
  and\ \bibinfo {author} {\bibfnamefont {M.}~\bibnamefont {Batchelor}},\
  }\href@noop {} {\bibfield  {journal} {\bibinfo  {journal} {J. Phys. A: Math.
  Gen.}\ }\textbf {\bibinfo {volume} {20}},\ \bibinfo {pages} {5677} (\bibinfo
  {year} {1987})}\BibitemShut {NoStop}%
\bibitem [{\citenamefont {Sirker}\ and\ \citenamefont {Bortz}(2006)}]{Sirker}%
  \BibitemOpen
  \bibfield  {author} {\bibinfo {author} {\bibfnamefont {J.}~\bibnamefont
  {Sirker}}\ and\ \bibinfo {author} {\bibfnamefont {M.}~\bibnamefont {Bortz}},\
  }\href@noop {} {\bibfield  {journal} {\bibinfo  {journal} {J. Stat. Mech.}\
  }\textbf {\bibinfo {volume} {2006}},\ \bibinfo {pages} {P01007} (\bibinfo
  {year} {2006})}\BibitemShut {NoStop}%
\bibitem [{\citenamefont {Morse}\ and\ \citenamefont {Feshbach}(1953)}]{Morse}%
  \BibitemOpen
  \bibfield  {author} {\bibinfo {author} {\bibfnamefont {P.~M.}\ \bibnamefont
  {Morse}}\ and\ \bibinfo {author} {\bibfnamefont {H.}~\bibnamefont
  {Feshbach}},\ }\href@noop {} {\emph {\bibinfo {title} {Methods of Theoretical
  Physics}}}\ (\bibinfo  {publisher} {New York, McGraw-Hill},\ \bibinfo {year}
  {1953})\BibitemShut {NoStop}%
\bibitem [{\citenamefont {Alcaraz}\ and\ \citenamefont
  {Wreszinski}(1990)}]{Alcaraz1990}%
  \BibitemOpen
  \bibfield  {author} {\bibinfo {author} {\bibfnamefont {F.~C.}\ \bibnamefont
  {Alcaraz}}\ and\ \bibinfo {author} {\bibfnamefont {W.~F.}\ \bibnamefont
  {Wreszinski}},\ }\href {\doibase 10.1007/BF01020284} {\bibfield  {journal}
  {\bibinfo  {journal} {J. Stat. Phys.}\ }\textbf {\bibinfo {volume} {58}},\
  \bibinfo {pages} {45} (\bibinfo {year} {1990})}\BibitemShut {NoStop}%
\bibitem [{\citenamefont {Yang}\ and\ \citenamefont
  {Yang}(1966{\natexlab{b}})}]{Yang-Yang}%
  \BibitemOpen
  \bibfield  {author} {\bibinfo {author} {\bibfnamefont {C.~N.}\ \bibnamefont
  {Yang}}\ and\ \bibinfo {author} {\bibfnamefont {C.~P.}\ \bibnamefont
  {Yang}},\ }\href@noop {} {\bibfield  {journal} {\bibinfo  {journal} {Phys.
  Rev.}\ }\textbf {\bibinfo {volume} {150}},\ \bibinfo {pages} {321} (\bibinfo
  {year} {1966}{\natexlab{b}})}\BibitemShut {NoStop}%
\bibitem [{\citenamefont {Affleck}\ and\ \citenamefont {Lieb}(1986)}]{Affleck}%
  \BibitemOpen
  \bibfield  {author} {\bibinfo {author} {\bibfnamefont {I.}~\bibnamefont
  {Affleck}}\ and\ \bibinfo {author} {\bibfnamefont {E.~H.}\ \bibnamefont
  {Lieb}},\ }\href@noop {} {\bibfield  {journal} {\bibinfo  {journal} {Lett.
  Math. Phys.}\ }\textbf {\bibinfo {volume} {12}},\ \bibinfo {pages} {57}
  (\bibinfo {year} {1986})}\BibitemShut {NoStop}%
\bibitem [{Note1()}]{Note1}%
  \BibitemOpen
  \bibinfo {note} {Since in Ref.~\cite {tanikawa2021exact} all the exponents
  $4m\gamma /(\pi -\gamma )~(m\in \protect \mathbb {N})$ are supposed to be
  noninteger, the points included in {$\protect \mathcal {S}_{e}$} or
  {$\protect \mathcal {S}_{l}$} are automatically excluded.}\BibitemShut
  {Stop}%
\bibitem [{Note2()}]{Note2}%
  \BibitemOpen
  \bibinfo {note} {The result for the linear Drude weight $(k=1)$ is consistent
  with Eq.~(20) in Ref.~\cite {laflorencie2001}, with the identification
  $K^*=\pi /(2 (\pi -\gamma ))$}\BibitemShut {NoStop}%
\bibitem [{Note3()}]{Note3}%
  \BibitemOpen
  \bibinfo {note} {We note in passing that a divergent behavior similar to that
  of $\protect \mathcal {D}^{(3)}_N(0)$ was found for the fourth derivative of
  the ground state energy density with respect to the magnetization~\cite
  {Nomura}.}\BibitemShut {Stop}%
\bibitem [{\citenamefont {Lukyanov}(1998)}]{Lukyanov}%
  \BibitemOpen
  \bibfield  {author} {\bibinfo {author} {\bibfnamefont {S.}~\bibnamefont
  {Lukyanov}},\ }\href@noop {} {\bibfield  {journal} {\bibinfo  {journal}
  {Nucl. Phys. B}\ }\textbf {\bibinfo {volume} {522}},\ \bibinfo {pages} {533}
  (\bibinfo {year} {1998})}\BibitemShut {NoStop}%
\bibitem [{\citenamefont {Sutherland}\ and\ \citenamefont
  {Shastry}(1990)}]{Sutherland}%
  \BibitemOpen
  \bibfield  {author} {\bibinfo {author} {\bibfnamefont {B.}~\bibnamefont
  {Sutherland}}\ and\ \bibinfo {author} {\bibfnamefont {B.~S.}\ \bibnamefont
  {Shastry}},\ }\href@noop {} {\bibfield  {journal} {\bibinfo  {journal} {Phys.
  Rev. Lett.}\ }\textbf {\bibinfo {volume} {65}},\ \bibinfo {pages} {1833}
  (\bibinfo {year} {1990})}\BibitemShut {NoStop}%
\bibitem [{\citenamefont {Bortz}\ \emph {et~al.}(2009)\citenamefont {Bortz},
  \citenamefont {Karbach}, \citenamefont {Schneider},\ and\ \citenamefont
  {Eggert}}]{PhysRevB.79.245414}%
  \BibitemOpen
  \bibfield  {author} {\bibinfo {author} {\bibfnamefont {M.}~\bibnamefont
  {Bortz}}, \bibinfo {author} {\bibfnamefont {M.}~\bibnamefont {Karbach}},
  \bibinfo {author} {\bibfnamefont {I.}~\bibnamefont {Schneider}}, \ and\
  \bibinfo {author} {\bibfnamefont {S.}~\bibnamefont {Eggert}},\ }\href
  {\doibase 10.1103/PhysRevB.79.245414} {\bibfield  {journal} {\bibinfo
  {journal} {Phys. Rev. B}\ }\textbf {\bibinfo {volume} {79}},\ \bibinfo
  {pages} {245414} (\bibinfo {year} {2009})}\BibitemShut {NoStop}%
\bibitem [{\citenamefont {Liu}\ \emph {et~al.}(2021)\citenamefont {Liu},
  \citenamefont {Fuji},\ and\ \citenamefont {Watanabe}}]{liu2021bloch}%
  \BibitemOpen
  \bibfield  {author} {\bibinfo {author} {\bibfnamefont {Y.}~\bibnamefont
  {Liu}}, \bibinfo {author} {\bibfnamefont {Y.}~\bibnamefont {Fuji}}, \ and\
  \bibinfo {author} {\bibfnamefont {H.}~\bibnamefont {Watanabe}},\ }\href
  {\doibase 10.1103/PhysRevB.104.205115} {\bibfield  {journal} {\bibinfo
  {journal} {Phys. Rev. B}\ }\textbf {\bibinfo {volume} {104}},\ \bibinfo
  {pages} {205115} (\bibinfo {year} {2021})}\BibitemShut {NoStop}%
\bibitem [{\citenamefont {Deguchi}\ \emph {et~al.}(2001)\citenamefont
  {Deguchi}, \citenamefont {Fabricius},\ and\ \citenamefont
  {McCoy}}]{deguchi2001sl2}%
  \BibitemOpen
  \bibfield  {author} {\bibinfo {author} {\bibfnamefont {T.}~\bibnamefont
  {Deguchi}}, \bibinfo {author} {\bibfnamefont {K.}~\bibnamefont {Fabricius}},
  \ and\ \bibinfo {author} {\bibfnamefont {B.~M.}\ \bibnamefont {McCoy}},\
  }\href@noop {} {\bibfield  {journal} {\bibinfo  {journal} {J. Stat. Phys.}\
  }\textbf {\bibinfo {volume} {102}},\ \bibinfo {pages} {701} (\bibinfo {year}
  {2001})}\BibitemShut {NoStop}%
\bibitem [{\citenamefont {Miao}\ \emph {et~al.}(2021)\citenamefont {Miao},
  \citenamefont {Lamers},\ and\ \citenamefont {Pasquier}}]{miao2021q}%
  \BibitemOpen
  \bibfield  {author} {\bibinfo {author} {\bibfnamefont {Y.}~\bibnamefont
  {Miao}}, \bibinfo {author} {\bibfnamefont {J.}~\bibnamefont {Lamers}}, \ and\
  \bibinfo {author} {\bibfnamefont {V.}~\bibnamefont {Pasquier}},\ }\href@noop
  {} {\bibfield  {journal} {\bibinfo  {journal} {SciPost Physics}\ }\textbf
  {\bibinfo {volume} {11}},\ \bibinfo {pages} {067} (\bibinfo {year}
  {2021})}\BibitemShut {NoStop}%
\bibitem [{Note4()}]{Note4}%
  \BibitemOpen
  \bibinfo {note} {We also note that the number of coupled nonlinear integral
  equations arising in the thermodynamic Bethe ansatz becomes finite at these
  points~\cite {Takahashi,takahashi-suzuki}.}\BibitemShut {Stop}%
\bibitem [{\citenamefont {Yu}\ and\ \citenamefont
  {Fowler}(1992)}]{PhysRevB.46.14583}%
  \BibitemOpen
  \bibfield  {author} {\bibinfo {author} {\bibfnamefont {N.}~\bibnamefont
  {Yu}}\ and\ \bibinfo {author} {\bibfnamefont {M.}~\bibnamefont {Fowler}},\
  }\href {\doibase 10.1103/PhysRevB.46.14583} {\bibfield  {journal} {\bibinfo
  {journal} {Phys. Rev. B}\ }\textbf {\bibinfo {volume} {46}},\ \bibinfo
  {pages} {14583} (\bibinfo {year} {1992})}\BibitemShut {NoStop}%
\bibitem [{\citenamefont {Spivey}(2006)}]{Spivey}%
  \BibitemOpen
  \bibfield  {author} {\bibinfo {author} {\bibfnamefont {M.~Z.}\ \bibnamefont
  {Spivey}},\ }\href {http://www.jstor.org/stable/27642905} {\bibfield
  {journal} {\bibinfo  {journal} {Math. Mag.}\ }\textbf {\bibinfo {volume}
  {79}},\ \bibinfo {pages} {61} (\bibinfo {year} {2006})}\BibitemShut {NoStop}%
\bibitem [{\citenamefont {Graham}\ \emph {et~al.}(1994)\citenamefont {Graham},
  \citenamefont {Knuth},\ and\ \citenamefont
  {Patashnik}}]{GKP-Concrete-Math-2nd}%
  \BibitemOpen
  \bibfield  {author} {\bibinfo {author} {\bibfnamefont {R.~L.}\ \bibnamefont
  {Graham}}, \bibinfo {author} {\bibfnamefont {D.~E.}\ \bibnamefont {Knuth}}, \
  and\ \bibinfo {author} {\bibfnamefont {O.}~\bibnamefont {Patashnik}},\
  }\href@noop {} {\emph {\bibinfo {title} {Concrete Mathematics---A Foundation
  for Computer Science}}},\ \bibinfo {edition} {2nd}\ ed.\ (\bibinfo
  {publisher} {Addison-Wesley Publishing Company},\ \bibinfo {address}
  {Reading, MA},\ \bibinfo {year} {1994})\BibitemShut {NoStop}%
\bibitem [{\citenamefont {Woynarovich}\ and\ \citenamefont
  {Eckle}(1987)}]{Woynarovich}%
  \BibitemOpen
  \bibfield  {author} {\bibinfo {author} {\bibfnamefont {F.}~\bibnamefont
  {Woynarovich}}\ and\ \bibinfo {author} {\bibfnamefont {H.-P.}\ \bibnamefont
  {Eckle}},\ }\href@noop {} {\bibfield  {journal} {\bibinfo  {journal} {J.
  Phys. A: Math. Gen.}\ }\textbf {\bibinfo {volume} {20}},\ \bibinfo {pages}
  {L97} (\bibinfo {year} {1987})}\BibitemShut {NoStop}%
\bibitem [{\citenamefont {Eckle}(2019)}]{eckle2019models}%
  \BibitemOpen
  \bibfield  {author} {\bibinfo {author} {\bibfnamefont {H.-P.}\ \bibnamefont
  {Eckle}},\ }\href@noop {} {\emph {\bibinfo {title} {Models of Quantum Matter:
  A First Course on Integrability and the Bethe Ansatz}}}\ (\bibinfo
  {publisher} {Oxford University Press, USA},\ \bibinfo {year}
  {2019})\BibitemShut {NoStop}%
\bibitem [{\citenamefont {Granet}\ \emph {et~al.}(2018)\citenamefont {Granet},
  \citenamefont {Jacobsen},\ and\ \citenamefont {Saleur}}]{GRANET201896}%
  \BibitemOpen
  \bibfield  {author} {\bibinfo {author} {\bibfnamefont {E.}~\bibnamefont
  {Granet}}, \bibinfo {author} {\bibfnamefont {J.~L.}\ \bibnamefont
  {Jacobsen}}, \ and\ \bibinfo {author} {\bibfnamefont {H.}~\bibnamefont
  {Saleur}},\ }\href {\doibase https://doi.org/10.1016/j.nuclphysb.2018.06.001}
  {\bibfield  {journal} {\bibinfo  {journal} {Nucl. Phys. B}\ }\textbf
  {\bibinfo {volume} {934}},\ \bibinfo {pages} {96} (\bibinfo {year}
  {2018})}\BibitemShut {NoStop}%
\bibitem [{\citenamefont {Laflorencie}\ \emph {et~al.}(2001)\citenamefont
  {Laflorencie}, \citenamefont {Capponi},\ and\ \citenamefont
  {S{\o}rensen}}]{laflorencie2001}%
  \BibitemOpen
  \bibfield  {author} {\bibinfo {author} {\bibfnamefont {N.}~\bibnamefont
  {Laflorencie}}, \bibinfo {author} {\bibfnamefont {S.}~\bibnamefont
  {Capponi}}, \ and\ \bibinfo {author} {\bibfnamefont {E.~S.}\ \bibnamefont
  {S{\o}rensen}},\ }\href@noop {} {\bibfield  {journal} {\bibinfo  {journal}
  {Eur. Phys. J. B}\ }\textbf {\bibinfo {volume} {24}},\ \bibinfo {pages} {77}
  (\bibinfo {year} {2001})}\BibitemShut {NoStop}%
\bibitem [{\citenamefont {Aiba}\ and\ \citenamefont {Nomura}(2020)}]{Nomura}%
  \BibitemOpen
  \bibfield  {author} {\bibinfo {author} {\bibfnamefont {N.}~\bibnamefont
  {Aiba}}\ and\ \bibinfo {author} {\bibfnamefont {K.}~\bibnamefont {Nomura}},\
  }\href {\doibase 10.1103/PhysRevB.102.134435} {\bibfield  {journal} {\bibinfo
   {journal} {Phys. Rev. B}\ }\textbf {\bibinfo {volume} {102}},\ \bibinfo
  {pages} {134435} (\bibinfo {year} {2020})}\BibitemShut {NoStop}%
\bibitem [{\citenamefont {Takahashi}\ and\ \citenamefont
  {Suzuki}(1972)}]{takahashi-suzuki}%
  \BibitemOpen
  \bibfield  {author} {\bibinfo {author} {\bibfnamefont {M.}~\bibnamefont
  {Takahashi}}\ and\ \bibinfo {author} {\bibfnamefont {M.}~\bibnamefont
  {Suzuki}},\ }\href@noop {} {\bibfield  {journal} {\bibinfo  {journal} {Prog.
  Theor. Phys}\ }\textbf {\bibinfo {volume} {48}},\ \bibinfo {pages} {2187}
  (\bibinfo {year} {1972})}\BibitemShut {NoStop}%
\end{thebibliography}

\end{document}